\documentclass[prd,reprint,nofootinbib,notitlepage,aps,tightenlines,amsmath,amssymb,showpacs,superscriptaddress]{revtex4-1}
\usepackage[hyperindex,breaklinks]{hyperref}
\usepackage{graphicx,epstopdf}
\usepackage{tikz}
\usetikzlibrary{shapes,arrows}

\usepackage{mathtools}

\usepackage{subfigure}

\usepackage{natbib,bm}
\usepackage{slashed}    
\usepackage{hyperref}
\usepackage{breakurl}
\usepackage{multirow}
\usepackage{bbold}
\usepackage{listings}
\usepackage{booktabs}

\usepackage{soul}

\def\be{\begin{equation}}
\def\ee{\end{equation}}
\def\bea{\begin{eqnarray}}
\def\eea{\end{eqnarray}}
\def\ge{\mathrel{\raise.3ex\hbox{$>$\kern-.75em\lower1ex\hbox{$\sim$}}}}
\def\la{\mathrel{\raise.3ex\hbox{$<$\kern-.75em\lower1ex\hbox{$\sim$}}}}

\def\simgt{\mathrel{\raise.3ex\hbox{$>$\kern-.75em\lower1ex\hbox{$\sim$}}}}
\def\simlt{\mathrel{\raise.3ex\hbox{$<$\kern-.75em\lower1ex\hbox{$\sim$}}}}

\newcommand{\nc}{\newcommand}

\nc{\gone}{\bar g_{\pi NN}^{(1)}}
\nc{\gzero}{\bar g_{\pi NN}^{(0)}}
\nc{\al}{\alpha}
\nc{\ga}{\gamma}
\nc{\de}{\delta}
\nc{\ep}{\epsilon}
\nc{\ze}{\zeta}
\nc{\et}{\eta}
\nc{\ka}{\kappa}
\nc{\rh}{\rho}
\nc{\si}{\sigma}
\nc{\ta}{\tau}
\nc{\up}{\upsilon}
\nc{\ph}{\phi}
\nc{\ch}{\chi}
\nc{\ps}{\psi}
\nc{\om}{\omega}
\nc{\Ga}{\Gamma}
\nc{\De}{\Delta}
\nc{\La}{\Lambda}
\nc{\Si}{\Sigma}
\nc{\Up}{\Upsilon}
\nc{\Ph}{\Phi}
\nc{\Ps}{\Psi}
\nc{\Om}{\Omega}
\nc{\ptl}{\partial}
\nc{\del}{\nabla}
\nc{\ov}{\overline}
\nc{\us}{U(1)$_S$}

\mathchardef\mhyphen="2D

\def\beq{\begin{equation}}
\def\eeq{\end{equation}}
\def\bmat{\begin{displaymath}}
\def\emat{\end{displaymath}}
\def\bear{\begin{eqnarray}}
\def\eear{\end{eqnarray}}
\def\ba{\begin{eqnarray}}
\def\ea{\end{eqnarray}}
\def\bery{\begin{array}}
\def\ery{\end{array}}
\def\bit{\begin{itemize}}
\def\eit{\end{itemize}}
\def\ben{\begin{enumerate}}
\def\een{\end{enumerate}}
\def\btab{\begin{tabular}}
\def\etab{\end{tabular}}
\def\btbl{\begin{table}}
\def\etbl{\end{table}}
\def\bfig{\begin{figure}[htb]}
\def\efig{\end{figure}}
\def\bpic{\begin{picture}}
\def\epic{\end{picture}}

\DeclareMathOperator{\gauss}{gauss}

\def\ga{\mathrel{\raise.3ex\hbox{$>$\kern-.75em\lower1ex\hbox{$\sim$}}}}
\def\la{\mathrel{\raise.3ex\hbox{$<$\kern-.75em\lower1ex\hbox{$\sim$}}}}
\def\gappeq{\mathrel{\rlap {\raise.5ex\hbox{$>$}}
{\lower.5ex\hbox{$\sim$}}}}
\def\lappeq{\mathrel{\rlap{\raise.5ex\hbox{$<$}}
{\lower.5ex\hbox{$\sim$}}}}

\def\gyr{{\rm \, G\kern-0.125em yr}}
\def\mev{{\rm \, Me\kern-0.125em V}}
\def\gev{{\rm \, Ge\kern-0.125em V}}
\def\tev{{\rm \, Te\kern-0.125em V}}

\newcommand{\keV}{\ensuremath{\mathrm{keV}}}
\newcommand{\MeV}{\ensuremath{\mathrm{MeV}}}
\newcommand{\GeV}{\ensuremath{\mathrm{GeV}}}

\renewcommand{\sec}{\ensuremath{\mathrm{s}}}
\newcommand{\seconds}{\ensuremath{\mathrm{s}}}

\newcommand{\cm}{\ensuremath{\mathrm{cm}}}

\newcommand{\km}{\ensuremath{\mathrm{km}}}

\newcommand{\phiegal}{\Phi_{\nu,\mathrm{e.gal}}}
\newcommand{\phigal}{\Phi_{\nu,\mathrm{gal}}}

\AtBeginDocument{
\heavyrulewidth=.08em
\lightrulewidth=.05em
\cmidrulewidth=.03em
\belowrulesep=.65ex
\belowbottomsep=0pt
\aboverulesep=.4ex
\abovetopsep=0pt
\cmidrulesep=\doublerulesep
\cmidrulekern=.5em
\defaultaddspace=.5em
}

\begin{document}
\title{The neutrino-floor in the presence of dark radation}
\author{Marco Nikolic}
\affiliation{Institute of High Energy Physics, Austrian
     Academy of Sciences, Nikolsdorfergasse 18, 1050 Vienna,
     Austria}
\author{Suchita Kulkarni}
\affiliation{Institute of High Energy Physics, Austrian
     Academy of Sciences, Nikolsdorfergasse 18, 1050 Vienna,
     Austria}
\author{Josef Pradler}
\affiliation{Institute of High Energy Physics, Austrian
     Academy of Sciences, Nikolsdorfergasse 18, 1050 Vienna,
     Austria}
\begin{abstract}
\noindent 
 In this work we analyse the ultimate sensitivity of dark matter direct detection experiments, the ``neutrino-floor", in the presence of  anomalous sources of dark radiation in form of SM or semi-sterile neutrinos. This flux-component is assumed to be produced from dark matter decay. Since dark radiation may mimic dark matter signals, we perform our analysis based on likelihood statistics that allows to test the distinguishability between signals and backgrounds.
We show that the neutrino floor for xenon-based experiments may be lifted in the presence of extra dark radiation. In addition, we explore the testability of neutrino dark radiation from dark matter decay in direct detection experiments. Given the previous bounds from neutrino experiments, we find that xenon-based dark matter searches will not be able to probe new regions of the dark matter progenitor mass and lifetime parameter space when the decay products are SM neutrinos. In turn, if the decay instead happens to a fourth neutrino species with enhanced interactions to baryons, DR can either constitute  the dominant background or a discoverable signal in direct detection experiments.
\end{abstract}
\maketitle
\section{Introduction}

The neutral current-induced coherent neutrino-nucleus scattering process~\cite{Freedman:1973yd,Kopeliovich:1974mv} once inspired the conception of dark matter (DM) direct detection experiments~\cite{Drukier:1983gj,PhysRevLett.55.25,Goodman:1984dc}. Neutrinos with energies up to several hundred MeV elastically scatter on atomic nuclei with a cross section that is approximately enhanced by the squared number of neutrons, $N^2$. Similarly,  the spin-independent scattering of weakly interacting massive particles (WIMPs)  is enhanced by the square of the atomic number, $A^2$, boosting the prospects of observing an atomic recoil signal from DM in ultra-low background detectors with keV energy thresholds. Whereas DM has yet to be directly observed in the laboratory, the very process of coherent neutrino-nucleus scattering that once started the field, may also be the defining process in closing the window of opportunity in our direct searches for electroweak-scale DM. Neutrinos produced in the sun, in the atmosphere or in supernova explosions, among other sources, constitute a steady flux that cannot be shielded and, given enough observation time, detector volume, and detection sensitivity will eventually be seen as an irreducible background in DM direct detection experiments. This limits the ultimate sensitivity to discover DM of mass $m_\chi$ and nucleon cross section $\sigma_n$, and the combination of parameters where this occurs is conventionally referred to as the ``neutrino floor''~\cite{Strigari:2009bq,Billard:2013qya, Monroe:2007xp}.

The previous years have seen steady advances in increasing the sensitivity of direct detection experiments. Besides a tremendous effort that is underway and aims at developing and operating ultra-low threshold detectors---see~\cite{Battaglieri:2017aum} and references therein---the classical WIMP detectors have now gone beyond the ton-yr mark in exposure, reaching a sensitivity of $\sigma_n \simeq 10^{-47}\,\cm^2$  and better at a DM mass $m_\chi$ of several tens of GeV. Full exposure results have been reported from liquid xenon experiments LUX~\cite{Akerib:2016vxi}, PANDAX-II~\cite{Cui:2017nnn}, and XENON1T~\cite{Aprile:2018dbl}, followed by results from current liquid argon detectors DEAP-3600~\cite{Ajaj:2019imk} and DarkSide-50~\cite{Agnes:2018fwg}. The next generation in these experiments, XENONnT~\cite{Aprile:2015uzo} and LZ~\cite{Akerib:2018dfk} is already under construction and/or commissioning and they will be sensitive enough to see a small number of neutrino background events.
Finally, exposures of several hundred ton-yr may be achieved with the respective liquid xenon and argon detectors DARWIN~\cite{Aalbers:2016jon} and DarkSide-20k~\cite{Aalseth:2017fik}, and their reach in $(m_\chi,\sigma_n)$ will be limited by the neutrino floor. 

More than 40 years after its prediction, coherent neutrino-nucleus scattering has finally been observed by the COHERENT collaboration using accelerator-based neutrino beams~\cite{Akimov:2017ade,Akimov:2020pdx};  efforts to detect the process using reactors are underway~\cite{Angloher:2019flc,Hakenmuller:2019ecb}. These measurements provide valuable new insight into the interactions of Standard Model (SM) neutrinos with the constituents of atomic nuclei, thereby constraining non-standard interactions to quarks and the presence of new forces. 
 In the context of DM direct detection, the most important neutrino source is the sun and beyond-SM neutrino physics utilizing these fluxes has been explored in~\cite{Harnik:2012ni,Pospelov:2012gm}; for more recent works see~\cite{Billard:2014yka,Dutta:2017nht,Bertuzzo:2017tuf,AristizabalSierra:2017joc,Shoemaker:2018vii,Boehm:2018sux,Gonzalez-Garcia:2018dep,AristizabalSierra:2019ykk,Chao:2019pyh} and the review~\cite{Dutta:2019oaj}. A common theme in many of these studies is that new interactions of neutrinos will modify, and typically elevate the standard neutrino floor by within a factor of a few, when the new physics is subjected to complementary constraints.  

In this work, we consider a principal alternative option. Rather than modifying neutrino interactions \textit{per se}, we shall primarily study the presence of new neutrino fluxes and their influence on DM detectability. Concretely, we consider DM decay as a source of SM neutrinos; only in a second step we shall also consider the possibility of new interactions in an extended neutrino sector.  Substantial fluxes of these neutrinos will originate from DM decay within our own galaxy as well as globally, by the cosmological decay of DM.  Indeed, it is entirely possible that the Universe is filled in significant number with relativistic particles, {\it dark radiation} (DR), that may have escaped detection to date. Taking the present energy density in dark matter, $\rho_{\rm DM}$, as a calibration point, DR may contribute as much as several per cent, $\rho_{\rm DR} \lesssim 0.1 \rho_{\rm DM}$, while still being allowed by gravitational cosmological probes~\cite{Poulin:2016nat}. When compared to the present number (average energy) of  cosmic microwave background (CMB) photons, $n_{\rm CMB}$ ($\langle E_{\rm CMB} \rangle$), this implies that%
\begin{align}
\frac{ \langle E_{\rm DR} \rangle}{\langle E_{\rm CMB} \rangle}  \lesssim 500  
\frac {n_{\rm CMB}}{n_{\rm DR}}.
\end{align}
Hence, at the expense of having much less DR quanta than CMB photons $n_{\rm DR} \ll n_{\rm CMB}$, their typical energy may be significantly larger, $\langle E_{\rm DR} \rangle \gg \langle E_{\rm CMB} \rangle $. If the energy is in the several tens of MeV ballpark, DR neutrinos induce keV-scale nuclear recoils in direct detection experiments, altering the predictions of the neutrino floor.

In fact, the neutrino energy range between 15-100~MeV is of particular interest because it is a window of opportunity---framed by solar and atmospheric neutrino fluxes at the respective low- and high-energy ends---to search for the diffuse supernova neutrino background (DSNB)~\cite{Beacom:2010kk,Beacom:2003nk}. Its non-observation to-date puts a limit on the flux of electron antineutrinos $\phi(\bar\nu_e) < 3\,/\cm/\seconds$~\cite{Bays:2011si}, and an upper limit on a cosmological $\nu_e$ flux from DM decay in this window
has  been established with super-Kamiokande data in~\cite{PalomaresRuiz:2007ry}; see also~\cite{Garcia-Cely:2017oco}.%
\footnote{Limits on the $\nu +\bar \nu$ flux from MeV-mass DM annihilation have been derived in~\cite{PalomaresRuiz:2007eu}. Fluxes of new stable decay products, ``boosted DM'' from DM decay or annihilation have \textit{e.g.}~been considered in~\cite{Huang:2013xfa,Agashe:2014yua,Agashe:2014yua}.}
In this work, we shall consider that (a component of) DM decays into neutrinos $\nu$ but not anti-neutrinos~$\bar\nu$. This possibility has been analyzed in some generality in a previous work by some of us~\cite{Cui:2017ytb}, where constrains on the combination of lifetime and progenitor mass  from neutrino and direct detection experiments have been derived. In this work we explore in detail the consequences on the DM neutrino floor in the allowed regions of parameter space. 

The paper is organized as follows: in Sec.~\ref{sec:fluxes} we establish the principal DR fluxes from DM decay, in Sec.~\ref{sec:DR_at_DD} we introduce the neutrino-induced event rates at direct detection experiments and list the standard neutrino fluxes, in Sec.~\ref{sec:stats} we establish the statistical tools for quantifying the principal reach for DM or DR discovery. The main results are then presented in Sec.~\ref{sec:results} before concluding in Sec.~\ref{sec:conclusions}.

\section{Dark Radiation from DM decay}
\label{sec:fluxes}
We consider the possibility that non-thermal DR is made of neutrinos, either from the Standard Model (SM) or of a new type, that originate from the decay of an unstable DM progenitor~$X$. Due to cosmological constraints~\cite{ Poulin:2016nat}, we restrict ourselves to DR scenarios, where only a mass-fraction $f_X = 10\,\%$ of the total DM abundance%
\footnote{
The precise statement, at 95\%~C.L., is that either 3.8\% of all of DM could have decayed between recombination and today, or that $f_X/\tau_X< 6.3\times 10^{-3}\,{\rm Gyr}$ for lifetimes larger than the age of the Universe. For simplicity we take $f_X=0.1$ and arbitrary lifetime~\cite{ Poulin:2016nat} even if it implies that we occasionally slightly slip into the disfavored region.}
injects monochromatic neutrinos via two body decays of a lepton-number carrying Majoron-type scalar relic $X\to \nu\nu $; a description of such asymmetric model is provided in the original paper~\cite{Cui:2017ytb}.
 As we are principally interested in neutral current processes, the flavor evolution of injected neutrinos is of no relevance. 

The DR flux arriving at the Earth is by and large a combination of two components, the galactic flux $\phigal$ and the extra-galactic flux $\phiegal$. Assuming no directional sensitivity, for a DM particle $X$ with  lifetime $\tau_X$ and mass $m_X$ decaying within the Milky Way, the differential galactic flux is given by,
\begin{equation}
\frac{d\phigal}{dE_\nu}= \frac{N_\nu\, f_X}{ \tau_X m_X }\,\displaystyle{ e^{-\frac{t_0}{\tau_X}} } r_\odot\rho_\odot \langle{J_\mathrm{dec}}\rangle \delta(E_\nu-E_\mathrm{in}),
\label{eq:gal_flux}
\end{equation}
where $t_0=  13.787 \pm 0.020$~Gyr is the age of the universe, $N_\nu=2$ is the number of neutrinos in final state, $E_\mathrm{in}=m_X/2$ is the injection energy, $r_\odot=8.33$~kpc is the distance between the observer at the Earth and the galactic center, $\rho_\odot=0.3 \; \mathrm{GeV}/\mathrm{cm}^3$ is the local DM density and $\langle J_\mathrm{dec}\rangle \approx 2.19$ is the angular averaged J-factor obtained from an NFW profile~\cite{Navarro:1995iw}. Compared to $t_0$, Eq.~\eqref{eq:gal_flux} probes the amount of DM that is decaying ``today.''

In contrast, DR that arrives from cosmological distances probes the decaying DM fraction at a time $t_{\rm dec} \leq t_0$, or, in terms of redshift, at $z\geq 0$. Hence, the 
extra-galactic flux is assembled by contributions from all redshifts. To estimate this flux, remember that neutrinos arriving at the Earth with relativistic energy $E_\nu=\left|\vec{p}_\nu \right|$ were emitted with higher energy $E^\mathrm{em}_{\nu} = E_\nu(1+z)$. The energy-differential extra-galactic flux is then obtained via a redshift integral, which for monochromatic injection can be resolved~\cite{Cui:2017ytb},
\begin{widetext}
\begin{equation}
\frac{d\phiegal}{dE_\lambda}=   N_\nu   \frac{f_X \, \Omega_\mathrm{dm} \rho_\mathrm{crit}}{H_0 m_X \tau_X}  \frac{1}{\left|\vec{p}_\nu \right|} \frac{1}{\sqrt{\alpha^3 \Omega_M+ \Omega_\Lambda}} \displaystyle{e^{-\frac{ t(\alpha-1)}{\tau_X}}} \Theta(\alpha-1),
\label{eq:egal_flux}
\end{equation}
\end{widetext}
where $\alpha= E_\mathrm{in}/E_\nu \geq 1$. We set the density parameters $\Omega_\mathrm{dm}h^2= 0.12,\, \Omega_M=0.315,\, \Omega_\Lambda=1-\Omega_M$ consistent with the $\Lambda$CDM model of a flat Universe.  $H_0=100 h$~km/s/Mpc and $\rho_\mathrm{crit}=3H^2/(8 \pi G)$ are the Hubble parameter and critical density at the present time with $h= 0.674$~\cite{Aghanim:2018eyx}.%
\footnote{
   There is a well-known current discrepancy between the CMB-inferred value of $H_0$ and the distance ladder estimates from SNIa~\cite{Riess:2019cxk} (among other, more recent local measurements). Our results only depend mildly on the adopted value of $H_0$, but we note that decaying DM scenarios such as the one considered here can alleviate this tension~\cite{Pandey:2019plg,Vattis:2019efj}.
} 
Considering DM lifetimes such that its decay proceeds after matter-radiation equality, we are allowed neglect the radiation content of the Universe and may find an analytic expression for the lookback time~$t(z)$,
\begin{widetext}
\begin{equation}
\begin{aligned}
\label{eq:lookback}
t(z)= \int_z^\infty \frac{dz'}{(1+z')H(z')}= \frac{1}{3H_0\sqrt{\Omega_\Lambda}}\ln \left|\frac{\sqrt{1+(1+z)^3\frac{\Omega_M}{\Omega_\Lambda}}+1}{\sqrt{1+(1+z)^3\frac{\Omega_M}{\Omega_\Lambda}}-1}\right| \, ,
\end{aligned}
\end{equation}
\end{widetext}
where $H(z)=H_0\sqrt{(1+z)^3 \Omega_M+ \Omega_\Lambda}$ is the Hubble rate at redshift $z$. In this way, a connection between the elapsed time and the redshift is established. Equations~\eqref{eq:gal_flux} and ~\eqref{eq:egal_flux} are also applicable for any two body massive final states with replacement $E_{\nu} \rightarrow \sqrt{m_\nu^2+\left|\vec{p}_\nu\, \right|^2}$, where $m_\nu, p_\nu$ are the mass and momenta of the massive species~\cite{Cui:2017ytb}.

The total DR flux $\Phi_{X2\nu}=\Phi_{\nu,\textnormal{gal.}}+\Phi_{\nu,\textnormal{e.gal}}$, a sum of the galactic and extra-galactic component, is obtained by integration over~$E_\nu$. The  galactic flux reads,
\begin{equation}
\phigal= N_\nu   \frac{f_X}{\tau_X m_X }\displaystyle{e^{-\frac{t_0}{\tau_X}}} r_\odot\rho_\odot \langle{J_\mathrm{dec}(\theta)}\rangle.
\end{equation}
One may in fact also obtain a useful analytic expression for the extra-galactic flux for the two-body decay with relativistic final states (see App.~\ref{app:ana_diff_rate}),
\begin{widetext}
\begin{equation}
\begin{aligned}\label{eq:PhiEg}
\phiegal &=   N_\nu  \frac{f_X\, \Omega_\mathrm{dm} \rho_\mathrm{crit}}{m_X }   \left[1- \left(\frac{\sqrt{\Omega_M/\Omega_\Lambda +1}+1}{\sqrt{\Omega_M/\Omega_\Lambda +1}-1} \right)^{-\frac{1}{3 H_0 \sqrt{\Omega_\Lambda} \tau_X}} \right],%
\end{aligned}
\end{equation}
\end{widetext}
where $\rho_{\rm crit}$ is the Universe's critical energy density today. %
The extra-galactic flux asymptotically approaches maximum for lifetimes $\tau_X \lesssim t_0$, and can be parametrised as %
\begin{equation}
    \Phi^\mathrm{max}_{\nu,\, \mathrm{e.gal}}\simeq 1.5\times 10^5 \, \left( \frac{f_X}{0.1}\right) \left( \frac{50 \, \mathrm{MeV}}{m_X} \right) \, \mathrm{cm}^{-2}\, \mathrm{s}^{-1} .
\end{equation}

\begin{figure}[tb]
  \includegraphics[width=\columnwidth]{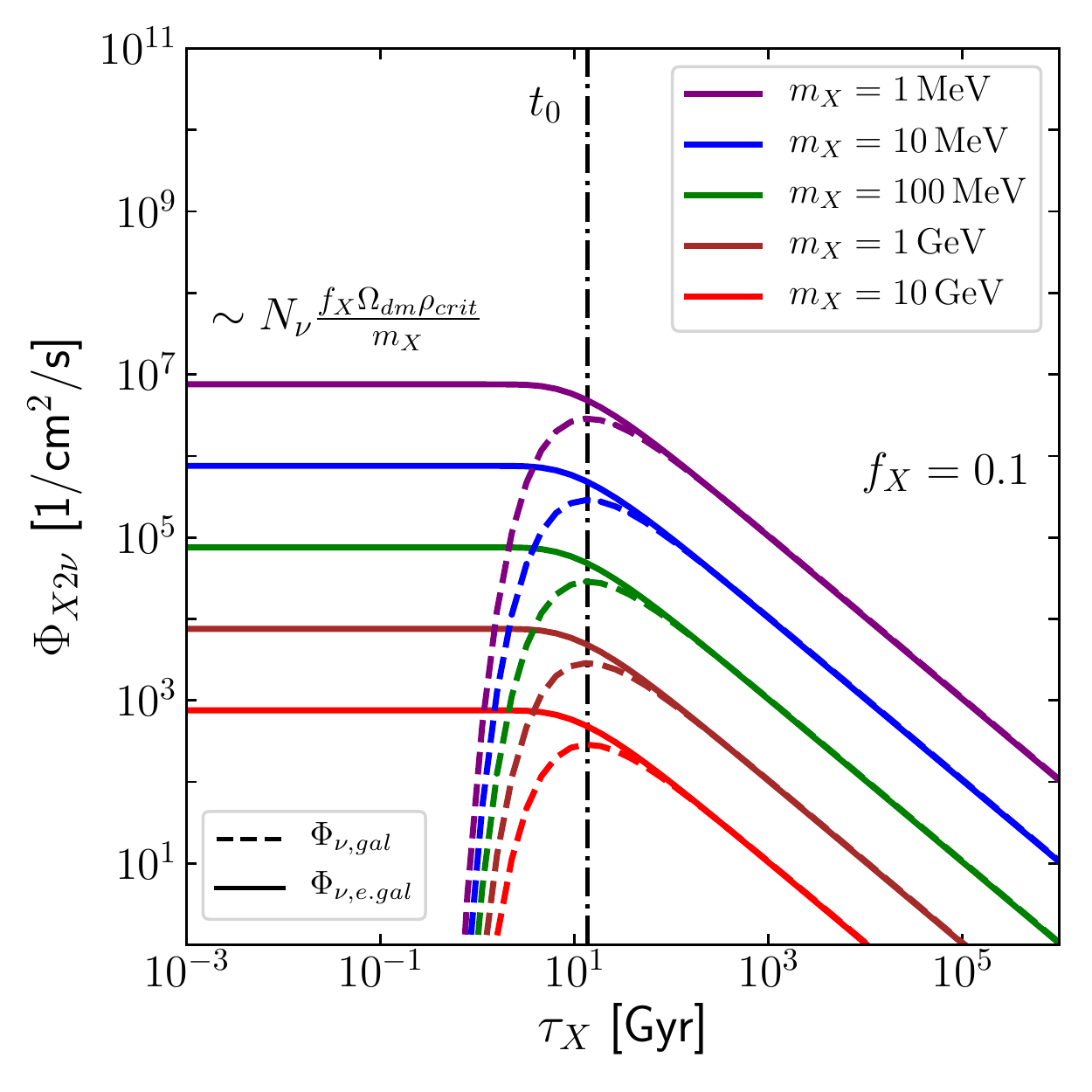}%
 \caption{The integrated neutrino flux originating from galactic (dashed lines) and extra-galactic (solid lines) components for different DM masses $m_X$ as a function of DM lifetime. The flux has the general $1/m_X$ scaling, is inversely proportional to progenitor lifetime for  $\tau_X\lesssim t_0$ and the extragalactic component asymptotes to a constant in the converse limit.  The maximum DR flux as the sum of solid and dashed lines is obtained for $\tau\sim t_0$. 
}
\label{fig:diff_flux}
\end{figure}

In Fig.~\ref{fig:diff_flux}, we exemplify the DR fluxes originating from galactic (dashed lines) and extra-galactic (solid lines) components for various progenitor masses $m_X$ assuming a 10\% mass-fraction of decaying DM, i.e.~$f_X=0.1$. As both fluxes are inversely proportional to progenitor mass, for a fixed progenitor lifetime the flux decreases as progenitor mass increases. For large lifetime, both galactic and extra-galactic fluxes fall exponentially, for small lifetime, the galactic flux diminishes exponentially, whereas the extra-galactic flux asymptotes to a constant value $N_{\nu}f_X\Omega_{DM}\rho_{crit}/m_X$. This is manifest from (\ref{eq:PhiEg}) and is easily understood: a {\it complete} decay of $X$ at some early redshift $z$ corresponds to a mere transferal from one particle species ($X$) to another ($2\nu$) with the comoving number of their sum being conserved.  ``Early DR'' originates from a cosmological DM density that is larger by a factor $(1+z)^3$, hence compensating for the flux dilution by a factor  $(1+z)^{-3} $ from area expansion and increase in the time-interval of subsequent particle arrivals. For  $\tau_X\ll t_0 $, the total extragalactic flux becomes hence independent of progenitor lifetime.
Nevertheless, given the finite energy thresholds of any experiment, the number of ``active'' neutrinos in a detector diminishes for $\nu$ originating at high redshift, because of their redshifting in energy. The maximum flux (as the sum of galactic and extragalactic contributions) is attained when $\tau_X\sim t_0$.
Finally, it is interesting to note that the galactic flux is a double-valued function in $m_X$: for a fixed progenitor mass, the  same flux is attained for two different progenitor lifetimes. As we will see later this feature leads to interesting consequences for the experimental sensitivity to DR.
\section{Signatures of DR at direct detection experiments}
\label{sec:DR_at_DD}
Dark radiation in the form of neutrinos produced by unstable progenitors as described in the previous section can potentially be detected at direct detection experiments at the Earth via neutrino-nucleus coherent scattering. 
Even in absence of such DR neutrinos, the neutrino-nucleus scattering at direct detection takes place due to neutrino fluxes arising from standard ambient neutrino sources. These include solar, atmospheric  and supernova-generated neutrinos. We shall refer to those fluxes as the ``standard'' ones. 

The standard neutrino fluxes adopted in this work together with their standard errors are listed in Tab.~\ref{tab:nuflux}. For solar neutrinos we use the ones based on the chemical composition determination of~\cite{Grevesse:1998bj}.%
\footnote{The results of this work are only mildly dependent on the solar opacity problem: the more recent determination~\cite{Asplund:2009fu}
  primarily affects the O, N, and F fluxes in addition to a $20\%$ downward shift of the ${}^8{\rm B}$ flux. We note in passing that for the last flux, being the most relevant in this context, the measurement~\cite{Vinyoles:2016djt} lies in between both low and metallicity determinations of~\cite{Asplund:2009fu} and~\cite{Grevesse:1998bj}, respectively.}  While solar neutrinos dominate for $E_\nu \lesssim 20 \, \rm MeV$, atmospheric and supernova neutrinos are otherwise the most important fluxes. There are no measurements of the atmospheric flux  below $100 \, \rm MeV$ and  we use the results from a FLUKA simulation~\cite{BATTISTONI2005526}.  The DSNB neutrinos originate from Type~II supernovae and their largest emission in all flavours takes place during the Kelvin-Helmholtz cooling phase. The prediction depends crucially on the star formation rate~\cite{Totani:1995dw} as a function of redshift and in this work we use the analytical fit provided in~\cite{Yuksel:2008cu}. The emission spectrum is expected to be thermal~\cite{HartmannD.H1997Tcsn}, with each neutrino component being assigned a specific temperature, $T_{\nu_e} \approx 4 \, $MeV, $T_{\bar{\nu}_e} \approx 5 \, $MeV, $T_{x} \approx 8 \, $MeV ($x=\nu_\mu,\bar{\nu}_\mu ,\nu_\tau ,\bar{\nu}_\tau $).
Finally, we note in passing that reactor- and geo-neutrinos are location dependent and relatively small, and we neglect them in this work.

\begin{table*}
\centering
\begin{tabular}{p{2.5cm}p{2.5cm}p{4cm}p{2.5cm}p{2.5cm}}
\toprule
Source	&				  & Flux $\Phi_\nu $ &  $E^\mathrm{max}_\nu$ &$E^\mathrm{max}_{R,\nu}$  \\
			 		 & &[cm$^{-2}$\,s$^{-1}$]  &$[$MeV$]$ &$[$keV$]$ \\ \midrule
solar &pp                  & $5.98\,(1 \pm 0.006)\times10^{10}$&$0.42$ & $2.9 \times 10^{-3}$ \\
&hep                 & $ 8.04\,(1 \pm 0.3)\times10^{3} $ & $1.88$ & $5.8$ \\
&$^{17}$F              & $5.52\,(1 \pm 0.17) \times 10^6$& $1.74$ & $5.0 \times 10^{-2}$ \\
&$^{15}$O              & $ 2.23\,(1 \pm 0.15)\times 10^8$ &$ 1.73$ & $4.9 \times 10^{-2}$ \\
&$^{13}$N              & $ 2.96\,(1 \pm 0.14)\times 10^8$ & $1.20$& $2.4 \times 10^{-2}$ \\
&$^8$B                 & $5.58\,(1 \pm 0.14)\times 10^6$ &$16.6$& $4.5$ \\
&$^7$Be (line 1)  & $4.50\,(1 \pm 0.07)\times 10^8$ &$0.384$ & $2.4 \times 10^{-3}$ \\
&$^7$Be (line 2) & $5.00\,(1 \pm 0.07)\times 10^9$ &$0.861$ & $1.2 \times 10^{-2}$ \\
&pep (line)   & $1.44\,(1 \pm 0.012)\times 10^8 $&$1.44$  & $3.4 \times 10^{-2}$ \\ \midrule
atmospheric & $\nu_e$          & $1.27\,(1 \pm 0.2)$ &$ 10^3$ & $>100$ \\
& $\bar{ \nu}_e$   & $1.17\,(1 \pm 0.2)$ &$ 10^3 $  & $> 100$ \\
& $\nu_\mu$        & $ 2.46\,(1 \pm 0.2)$ &$ 10^3$ & $> 100$ \\
& $\bar{ \nu}_\mu$ & $2.45\,(1 \pm 0.2)$ &$ 10^3 $ & $> 100$ \\ \midrule
DSNB & $\nu_e $ (fid)  & $22.12\,(1 \pm 0.5)$ &$  100$  & $> 100$ \\
&  $\overline{\nu}_e$ (fid)        & $ 17.69\,(1 \pm 0.5)$ &$ 100$ & $> 100 $ \\
& $\nu_x$ (fid)  & $11.06\,(1 \pm 0.5)$ &$ 100$ & $> 100 $ \\
\bottomrule
\end{tabular}
\caption{Adopted neutrino fluxes in this work; we follow~\cite{Gelmini:2018ogy,Battistoni:2005pd,Horiuchi:2008jz} in their compilation of the fluxes (original works are referened in the main text). Beside the fluxes, the endpoint or the maximally used energy together with the corresponding maximal nuclear recoil energies on a xenon target are given.
}
\label{tab:nuflux}
\end{table*}

The second relevant aspect is the differential recoil cross section of neutrinos on atomic nuclei. Within SM,  the process is mediated by neutral current interactions for spin-independent scattering and can be written as
\begin{equation}
\begin{aligned}
\label{eq:nu_nucleus_sigma}
\frac{d \, \sigma_{N \nu}(E_\nu, E_R)}{dE_R} &= \frac{Q_W^2G_F^2 \, m_N  F^2\left(\left|\vec{q}\,\right| \right)}{ 4 \pi } \left[1-\frac{ E_R m_N}{2E_\nu^2} \right],
\end{aligned}
\end{equation}
where $G_F=1.1663787(6) \times 10^{-5}$ GeV$^{-2}$ is the Fermi constant, $Q_W= (4 \sin^2 \theta_W -1) Z+N$ is the weak charge of the nucleus and $E_\nu$ is the neutrino energy. The differential xsec is primarily a function of number of Neutrons ($N$) because of a near cancellation in the charge ($Z$) dependent part of $Q_W$;  $\sin^2 \theta_W \approx 0.23$ is the weak angle. The degree of coherence is given by the  Helm form factor $F(|\vec{q}|)$~\cite{Lewin:1995rx} where $\vec q$ is the three-momentum transfer to the nucleus. 

As a second possibility we shall consider the case that $X$ decays into a pair of new neutrinos $\nu_B$ which interact with baryon number through a new vector particle of mass $m_V$ and gauge coupling $g_B$~\cite{Pospelov:2011ha}.
In the following we shall only be concerned with relativistic states and we may take the mass of the new neutrino to zero, $m_{\nu} \to 0$, so that $|\vec p_{\nu}| = E_{\nu}$.
The nuclear recoil cross section is coherently enhanced with atomic number $A^2$ and then reads~\cite{Pospelov:2011ha,Cui:2017ytb},
\begin{align}
  \label{eq:nub_nucleus_sigma}
   \frac{d\sigma_{N \nu}(E_\nu, E_R)}{dE_R} = \frac{A^2 q_\nu^2 g_B^4   }{2\pi}  \frac{  m_NF^2(|\vec q|) }{  (m_V^2 + 2m_N E_R)^2  }
   \left[ 1  - \frac{m_N E_R}{2 E_{\nu}^2  }\right] . 
\end{align}
For the sake of presentation, we shall only consider  the case of a mediator that is heavy compared to the typical momentum transfer $|\vec q| = \sqrt{2 E_R m_N}$, implying $m_V\gtrsim 100\,\MeV$ in practice. In that case the strength of the new interaction can parameterized by $G_B \equiv q_\nu g_B^2/m_V^2 $ allowing for a convenient comparison to $G_F$ of the SM sector. The phenomenology of this model has been explored in \cite{Pospelov:2013rha,Batell:2014yra,Dror:2017ehi,Dror:2017nsg,Aguilar-Arevalo:2018wea}, and in this work we shall consider allowed values $G_B>G_F$ as a possibility to boost the direct detection phenomenology of DR.

The differential recoil rate introduced by  neutrinos of source $i$ is therefore given by
\begin{equation}
\label{eq:SM_nu_recoil}
\frac{dR_{i}(E_{R})}{dE_R}  =N_T  \int_{E_{\nu,\mathrm{min}}}^{E_\nu^\mathrm{max}} \; dE_\nu \; \frac{d\Phi_{\nu,i}(E_\nu)}{dE_\nu} \frac{d \sigma_{N \nu}(E_\nu, E_R)}{dE_R}.
\end{equation}
where $E_{\nu,\mathrm{min}}= \sqrt{E_R m_N/2}$ is the minimum neutrino energy to produce a recoil on a target of mass $m_N$; $E_\nu^\mathrm{max}$ is the maximum energy neutrinos of source i can have. In case of DR, $E_\nu^\mathrm{max}$ is given by endpoint energy of the source $E_{in} = m_X /2$. The number of target nuclei per unit mass of detector material is denoted by $N_T$. The total rate in a detector will then be the sum over all  isotopic compositions of relevant elements and over all neutrino sources $i$. We shall denote by ${dR_{\nu}(E_{R})}/{dE_R} $ the total rate over all ``standard'' sources and by ${dR_{X2\nu}(E_{R})}/{dE_R}$ the DR-induced non-standard recoil rate. 
The latter carries two contributions, from galactic and extragalactic fluxes, respectively. An analytic expression for the differential recoil rate for the extragalactic flux can be obtained and is given in App.~\ref{app:ana_diff_rate}.

The maximum nuclear recoil energy from DR is derived from kinematics and depends on the mass of the progenitor, 
\begin{equation}
E^\mathrm{max}_{R,X2\nu} =\frac{2(E^\mathrm{max}_\nu)^2}{2 E^\mathrm{max}_\nu+m_N }
\simeq  \frac{m_X^2}{2m_N}  \quad (m_X \ll m_N) ,
\end{equation}
where $E^\mathrm{max}_\nu= E_{\mathrm{in}}$ is the neutrino energy at injection. It is worth noticing that the end point of recoil energy is directly proportional to the square of progenitor mass. 
The expected recoil rate is then given by the integral over the recoil energy,
\begin{equation}
\label{eq:rate_X2nu}
\mu_{X2\nu} =  \sum_{i=\textnormal{gal,e.gal}} \int_{E_{R_\mathrm{thr}}}^{E^\mathrm{max}_{R,X2\nu}} dE_R \; \frac{dR_i}{dE_R},
\end{equation}
where $E_{R_\mathrm{thr}}$ is the threshold of the detector. 

As an example, consider the recoil rate induced by a progenitor of 100~MeV mass decaying to SM neutrinos in a liquid xenon detector with negligible nominal 1~eV threshold.
Notwithstanding a general dependence on recoil energy, saturating the possible DR fluxes, we find that the differential rate is bounded from above by
\begin{equation*}
    \frac{dR_{X2\nu}}{dE_R}\lesssim
     \left( \frac{100\,\MeV}{m_X} \right)   
     \begin{cases}
    11\,\, \mathrm{keV}^{-1} \, \mathrm{ton}^{-1}  \, \mathrm{yr}^{-1} & \text{(gal.)} \\ 18 \,\,\mathrm{keV}^{-1} \, \mathrm{ton}^{-1}  \, \mathrm{yr}^{-1} & \text{(e.g.)} 
    \end{cases}.
\end{equation*}
This implies for the total detectable rate,
\begin{equation*}
    \mu_{X2\nu} \lesssim  \left( \frac{100\,\MeV}{m_X} \right) 
    \begin{cases}
    80\,\,  \mathrm{ton}^{-1}  \, \mathrm{yr}^{-1} & \text{(gal.)} \\
    124\,\,  \mathrm{ton}^{-1}  \, \mathrm{yr}^{-1} & \text{(e.g.)}
    \end{cases},
\end{equation*}
 and demonstrates that for a given progenitor mass, there is a natural ceiling on the event rate when considering DR in form of SM neutrinos. 

 \begin{figure}[tb] \includegraphics[width=0.97\columnwidth]{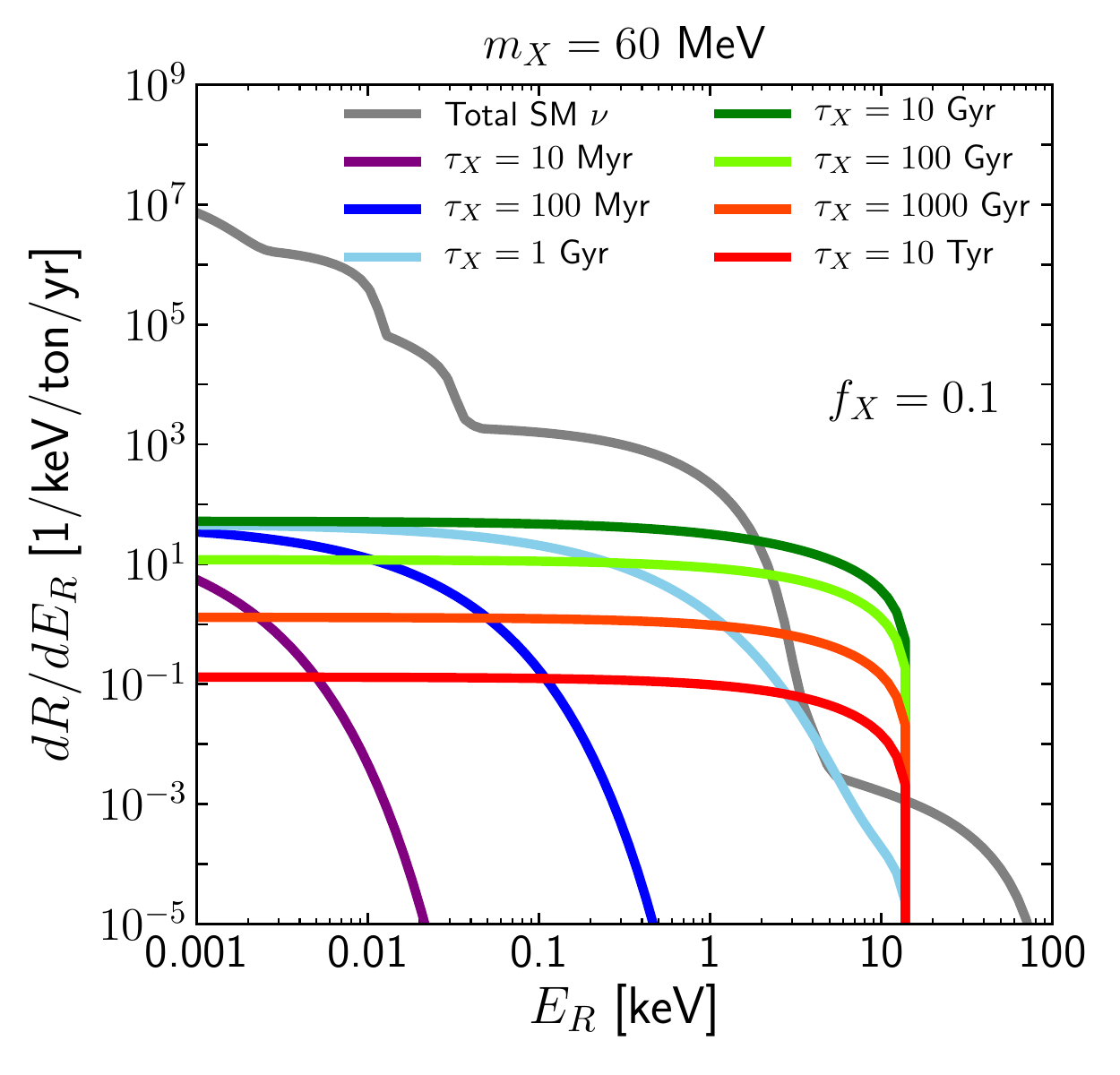}
   \caption{Energy differential recoil rate on xenon induced by standard fluxes (gray line) and induced by DR originating from a progenitor with $m_X =60\,\MeV$ and various lifetimes as labeled (colored lines).}
\label{fig:diff_recoil_rate}
\end{figure}
In Fig.~\ref{fig:diff_recoil_rate}, we show  exemplary differential recoil rates in xenon introduced by DR in form of SM neutrinos for a progenitor mass of 60 MeV in comparison with the differential recoil rate due to standard neutrino fluxes. The progenitor lifetime takes on values from $10^{-3}$~Gyr to $10^4$~Gyr. Heavier progenitor particles introduce larger recoil energies at direct detection experiments,  as the recoil endpoint energy is controlled by~$m_X$. Generally speaking, $m_X\gtrsim 30\,\MeV$ is required to exceed solar neutrino-induced events---in particular the ones induced by ${}^8$B neutrinos---because of the principal limitations in the magnitude of incident DR flux.

The latter point is further exacerbated by the fact that for a percent-level decaying fraction of DM we find that there is \textit{no} combination of $(\tau_X,m_X)$ such that the total DR-induced event rate in xenon is larger than the one from ${}^{8}$B neutrinos, 
\begin{align}
  R({\rm DR}) < R({\rm {}^{8}B})  \quad (E_{R_{\rm thr}}  \ll 1 \,\keV,\, m_X > 1\,\MeV ).
\end{align}
While it is well known that approximately 6~GeV mass WIMP recoils mimic  $^8$B neutrino recoils, we find that DR induced neutrino recoils from 2-body DM decay can neither mimic nor exceed the latter. This puts a principal limitation on the detectability/influence of DR in direct detection experiments: any set of prospective parameters must be such that the DR flux has a component that exceeds the  ${}^{8}$B flux in energy, hence pointing to $m_X\gtrsim 30\,\MeV$.%
\footnote{More precisely, if the statistical error bar on the ${}^{8}$B-induced events becomes smaller than the DR-induced event rate, one may actually subtract the ${}^{8}$B background and DR becomes detectable. This is seen by the extra island at $m_X\simeq 10\,\MeV$ in left Fig.~\ref{fig:discovery_contour_g} for the futuristic exposure of 100~ton-yr (see below).}

We now proceed and introduce the formalism to make quantitative statements on sensitivity of direct
detection experiments to a WIMP signal in presence of DR and/or to the
detecability of DR itself.

\section{The principle reach of direct detection experiments}
\label{sec:stats}

To forecast sensitivity of an experiment, there are two seemingly similar, yet distinct questions to address: the first regards the ability to {\rm exclude} the presence of a signal and the second regards the ability to {\rm discover} the signal. The exclusion and discovery exercises can in turn be performed either assuming  no backgrounds or in the presence thereof.

The purpose of this work is to quantify the reach of direct detection experiments to a DM signal in  presence of DR in addition to the standard neutrino background. Furthermore, we like to see to what extent direct detection experiments can discover DR.
We therefore deal with two different signal and background hypotheses, one which involves a DM signal and DR plus standard neutrino background, and one which involves DR recoils as signal and standard neutrino sources as background. In each case, we are interested in understanding the exclusion potential and discovery reach. 

Before discussing the statistical criteria, we first review the DM ($\chi$) signal hypothesis, which has not been discussed so far. The total event rate of DM recoils at the direct detection experiments, integrated over the differential recoil spectrum $dR_\chi/dE_R$ reads
\begin{widetext}
\begin{equation}
\begin{aligned}
\label{eq:WIMPevent}
\mu_\chi= \int_{E_{\mathrm{thr}}} \frac{dR_{\chi}}{dE_R} \, dE_R =  N_T \sigma_n  \displaystyle\frac{\rho_\chi}{m_\chi} \frac{m_N}{2 m_{\chi,n}^2}A^2\int_{E_{\mathrm{thr}}} \, dE_R\, F^2(|\vec{q}\,|)   \int_{|\vec{v}\,|\geq v_{\mathrm{min}}} d^3\vec{v}\, \displaystyle\frac{f(\vec{v}\,)}{v} 
\end{aligned}
\end{equation}
\end{widetext}
In the second equality, we make the assumption of a standard spin-independent DM-nucleus interaction of contact type. The local DM mass density is  $\rho_{\chi} \simeq 0.3\,\GeV/\cm^3$ and $m_\chi$ ($\mu_{\chi,n}$) is the DM (reduced DM-nucleon) mass  and $\sigma_n$ is the DM-nucleon scattering cross-section with equal coupling to protons and neutrons. At last, $f(\vec{v}\,)$ is the normalized galactic Maxwell-Boltzmann velocity distribution with $v_{0} = 220\,\km/\sec$, truncated at the escape velocity $v_{\rm esc} = 544\,\km/\sec$ and boosted into the detector frame with $v_{\rm lab} = 232\,\km/\sec$. Detector specifics are only entering via the recoil threshold energy $E_{\rm thr}$ and assume 100\% efficiency of detection for $E_R \geq E_{\rm thr}$, ignoring finite energy resolution effects. Furthermore, for our purposes it is sufficient to integrate the differential recoil rate up to a recoil energy of 100~keV. It should be noted that $\mu_{\chi}, \mu_{\nu},$ and $ \mu_{X2\nu}$ represent recoil rates per detector mass and live-time and need to be multiplied by detector exposure $\varepsilon$ to get the total number of observed events. 
\subsection{Exclusion potential of direct detection experiments}
\label{sec:stats_exclusion}
In this part, we set up the formalism to obtain the exclusion potential of direct detection experiments, for WIMP signal or for DR signal-only hypothesis. For this we assume the limiting case of zero background and zero observed signal events. It is clear, however, that for arbitrary large exposures, neutrino-induced events unavoidably lead to backgrounds and this limit will no longer be valid.

For a single channel counting experiment it is possible to take the number of events $n$ as the test statistic. The probability to observe $n$ events, when $\lambda$ events are expected is given by the Poisson distribution
  $   \mathrm{Pois}(n| \lambda)= \displaystyle {\lambda^{n} \, e^{-\lambda}}/{n!}.$
Here $\lambda$ can be the expected number of signal events $\varepsilon \mu_\chi$ or $\varepsilon \mu_{X2\nu}$, background events, or their sum. 
For zero observed events, $n=0$, and zero expected background, the 90\%~C.L.~upper limit on $\lambda$ is obtained from a p-value $p_\lambda=0.1$, corresponding to a 10\%\ probability that the outcome of an experiment is at least as extreme as observed. It is then related to the confidence level via $p_{\lambda}\leq 1-$CL. Solving $p_\lambda = \mathrm{Pois}(0| \lambda)$  for $\lambda$ yields $\lambda=2.3$ as usual.

In a background-free experiment with threshold energy $E_{\rm thr}$ and exposure $\varepsilon$, the $90\%$~C.L. exclusion contour in the ($\sigma_n$, $m_\chi$)-plane is found by  evaluating~\eqref{eq:WIMPevent} under the condition $\varepsilon \mu_\chi = 2.3$. The best exclusion that can be obtained in such background-free scenario, i.e.~the smallest value of $\sigma_n$ that can be excluded for a given DM mass $m_\chi$, is the one where the exposure grows to the level that the first irreducible neutrino background event is seen,
\begin{equation}
\label{eq:exp1nu}
    1 \stackrel{!}{=}\varepsilon(E_{\mathrm{thr}})\,\mu_\nu=\varepsilon(E_{\mathrm{thr}}) \int_{E_{\mathrm{thr}}} dE_R \, \displaystyle\frac{dR_\nu}{dE_R} .
\end{equation}
Here we have highlighted the threshold-dependence of the required exposure, $ \varepsilon(E_{\mathrm{thr}})$; the neutrino-induced recoil rate is given by the sum over all ``standard sources'' in Eq.~\eqref{eq:SM_nu_recoil}. Using (\ref{eq:exp1nu}) to find the minimum value of $\sigma_n$ that can be probed in a background-free experiment is at times colloquially referred to as the ``neutrino-floor''; we will, however, reserve the term for the discovery reach introduced later and not for the  exclusion boundary.
The presence of DR induced recoils leads to an additional source of background and hence needs to be accounted for while scaling the exposure. Hence the condition now reads %
\begin{equation}
    1 \stackrel{!}{=}
    \varepsilon(E_{\mathrm{thr}}) \left[\mu_\nu + \mu_{X2\nu} \right],
\end{equation}
in an obvious modification to Eq.~(\ref{eq:exp1nu}); the DR neutrino induced recoil rate is denoted by $ \mu_{X2\nu}$. The DR rate $\mu_{X2\nu}$ is a function of the progenitor mass and lifetime, we will take combinations of $(m_X, \tau_X)$  to evaluate the exclusion potential.

Finally, it is also possible to consider DR as a signal instead of background and we can estimate the potential of direct detection experiments to set constraints on the DR progenitor parameters $(m_X,\tau_X)$.  Assuming zero standard neutrino backgrounds, a signal is excluded at 90\%~C.L.~once $ \varepsilon  \mu_{X2\nu} \geq 2.3$ where $\mu_{X2\nu}$ is the DR-induced total event rate given in Eq.~(\ref{eq:rate_X2nu}). In exact analogy to the WIMP case, the exclusion potential for DR is found fixing the exposure to the value for which standard neutrino sources start being seen, Eq.~(\ref{eq:exp1nu}). In this case, however, $\tau_X$ and $m_X$ appear inside of the integral of (\ref{eq:SM_nu_recoil}), and the contour  $ \varepsilon  \mu_{X2\nu} = 2.3$ in the $(m_X,\tau_X)$ plane needs to be extracted numerically.

\subsection{Discovery potential of direct detection experiments}
\label{sec:stats_discovery}
While the procedure described above works well in the zero background scenario, the upcoming/ongoing ton scale direct detection experiments with large exposures will not remain background free. These backgrounds which originate from standard neutrino recoils suffer from uncertainties in the measured neutrino fluxes (see Tab.~\ref{tab:nuflux}) and therefore need to be modelled as distributions rather than fixed numbers. Hence, the statistical procedure  needs to take  these background distributions as nuisance parameters into account. This is accomplished by means of a profile likelihood analysis.

We now proceed and estimate the discovery potentials for the upcoming direct detection experiments and  begin by introducing the likelihood functions necessary for the profile likelihood analysis. Any observed spectrum of events is composed of signal and background sources. They all enter the generalized Poisson likelihood,
\begin{align}
  \label{eq:Levents}
  \mathcal{L}_{\text{events}} (\vec \theta | {\rm H}) = \frac{e^{- \varepsilon  \sum_{\alpha} \mu_\alpha (\vec \theta)}}{N_{\rm obs}!}  \prod_{i=1}^{N_{\rm obs}}  \varepsilon \sum_{
  \alpha} \frac{dR_\alpha(E_{R_i}, \vec \theta)}{dE_R}  ,
\end{align}
where the sum over sources $\alpha$ depends on the hypothesis under question. In this work we consider DM, DR, and standard neutrino-induced events, and $\alpha = \mathrm{DM}, X2\nu, \text{and }\nu_j$ together with their associated recoil spectra $ dR_\alpha(E_{R_i}, \vec \theta)/dE_R $ are possible. Model parameters are part of the vector $\vec\theta$ and for DM they are $m_{\chi}$ and $\sigma_n$, for DR they are $m_X$ and $\tau_X$  and for $\nu_j$ they are the assumed standard fluxes $\phi_{\nu}^j$.   In our simulations, the individual recoil events $E_{R_i}$ are random numbers. A total of $N_{\alpha} $ of them  for each source $\alpha$ are drawn from the probability density function (PDF) $ \mu_{\alpha}^{-1} dR_\alpha/dE_R$. Note that $N_{\alpha} $ itself is a Poisson random number with mean value  $\epsilon \mu_{\alpha}$. 
Finally, $N_{\rm obs} =\sum_{\alpha} N_{\alpha}$ is the total number of observed events.

When the number of observed events becomes large, one may also switch to a binned version of Eq.~(\ref{eq:Levents}),
\begin{align}
  \label{eq:LeventsBin}
  \mathcal{L}_{\text{events}}^{\text{binned}} (\vec \theta | {\rm H}) = \prod_{i=1}^{N_{\rm bin}}  \frac{e^{- \varepsilon  \sum_{\alpha} \mu_{\alpha}^i (\vec \theta)}  }{N_{\rm obs}^i!} \, \left(\sum_{\alpha} \mu_{\alpha}^i (\vec \theta) \right)^{N_{\rm obs}^i}. 
\end{align}
Here $N_{\rm bin}$ is the number of considered bins, $\mu_{\alpha}^i$ is the expected number of events for $\alpha=X2\nu, \nu_j$ and $N_{\rm obs}^i= \sum_\alpha N_\alpha ^i$ is the total number of observed events in each bin~$i$; in practice,  $N_{\rm obs}^i$ is the random number that needs to be drawn. Unless otherwise stated, we shall use the unbinned version~\eqref{eq:Levents} below.

When a source of events is declared a background under the considered hypothesis, we additionally subject their fluxes to appropriate uncertainties. Concretely, we model them as Gaussians with mean $\langle \phi_{\beta} \rangle $  and variance~$\sigma^2_{\beta}$,
\begin{align}
  \label{eq:Lflux}
  \mathcal{L}_{\text{bg-flux}}(\vec\theta) =  \prod_{\beta}  \gauss (\phi_{\beta} |  \langle \phi_{\beta} \rangle ,  \sigma^2_{\beta}) . 
\end{align}
Here, the product is over  all background sources  $\beta$ and the random variables $  \phi_{\beta} $ are  part of $\vec \theta$. 
For the standard neutrino sources the mean values $ \langle \phi_{\beta} \rangle  $ together with their standard errors are listed in Tab.~\ref{tab:nuflux}. When DR is considered as a background, we take $\langle \phi_{X2\nu} \rangle = \phiegal+\phigal$ and assume $\sigma_{X2\nu}=0.3$. In addition, we rewrite the differential rate as $dR_{ X2\nu}/dE_\nu = \phi_{X2\nu} \times \langle \phi_{X2\nu} \rangle^{-1}\langle dR_{ X2\nu}/dE_\nu  \rangle$. Here $\langle \, . \,\rangle$ denotes quantities evaluated at fixed values $(\tau_X,m_X)$. The assignment of a 30\%\ uncertainty  is somewhat arbitrary, but meant to be conservatively reflective of overall astrophysical uncertainties concerning DM-induced DR fluxes. Finally, DM is never considered as a background in this work and its flux is not part of (\ref{eq:Lflux}). Taken together,  the total likelihood becomes,
\begin{align}
  \label{eq:L}
  \mathcal{L}( \vec \theta | {\rm H}) =  \mathcal{L}_{\text{events}} (\vec \theta | {\rm H}) \times \mathcal{L}_{\text{bg-flux}}(\vec\theta) 
\end{align}
For a given threshold energy $E_{\rm thr}$ and exposure $\varepsilon$ and for every point $\vec \theta$ in the parameter space, we then generate several hundred mock realizations to build a statistical sample. Each of these realisations we refer to as an ``experiment". 

For each experiment one may then use the negative log-likelihood as test statistic~\cite{cowan1998statistical},
\begin{equation}
\label{eq:test_statistic}
q=-2 \ln \frac{ \mathcal{L}( \hat{\hat{\vec \theta}} | {\rm H_0}) }{ \mathcal{L}( \hat {\vec \theta} | {\rm H_1})},
\end{equation}
where $\hat{\hat{\vec \theta}}$ maximizes the likelihood under the background-only hypothesis H$_0$ and $\hat {\vec \theta}$ maximizes $\mathcal{L}$ for  signal plus background, H$_1$.  The distribution of $q$ under H$_0$ asymptotically follows a $\chi^2$-distribution with one degree of freedom  as per Wilk's theorem~\cite{wilks1938}. 
The $p_0$-value then characterizes the probability that the background-only hypothesis H$_0$ is excluded if the $q$-value is greater than the ``observed value'' $q_\mathrm{obs}$,
\begin{equation}
p_0= \mathrm{P} (q \geq q_\mathrm{obs}|\, \mathrm{H}_0)= \int_{q_\mathrm{obs}}^\infty dq \, f(q | \, \mathrm{H}_0) ,
\end{equation}
where $f(q | \, \mathrm{H}_0)$ is the PDF of $q$ under $\mathrm{H}_0$. We obtain the latter by Monte Carlo generation of mock data as described in the next section.  In terms of significance  $Z=\sqrt{q}$, given in units of standard deviations, the $p_0$-value is obtained by the cumulative standard normal distribution $\Phi(x)$,
\begin{equation}
p_0= \mathrm{P} (Z \geq Z_\mathrm{obs}|\, \mathrm{H}_0)\simeq
1- \Phi(Z_\mathrm{obs}).
\end{equation}
 For a discovery with $3 \sigma$ significance the corresponding $p_0$-value is given by $p_0$ = 0.00135. If data is observed in the critical region $Z \geq Z_{\rm obs}$ the hypothesis $H_0$ is rejected. For the $3 \sigma$ significance that we are interested in, $Z_{\rm obs} = 3$. We have verified, that our generated background data satisfies $f(q|\,\mathrm{H}_0) \sim \chi^2$.

In turn, the probability $\beta$ for $H_1$ being rejected is defined by,
\begin{equation}
\label{eq:H1rejection}
\beta = \mathrm{P} (Z \leq Z_{90}|\, \mathrm{H}_1)= \int_{0}^{Z_{90}} dZ \, f(Z | \, \mathrm{H}_1),
\end{equation}
where $f$ is now the PDF of $Z$ under $H_1$ and
$Z_{90}$ represents the significance that can be at least obtained $90\%$ of the time in an experiment with data generated under $H_1$. Conversely, H$_1$ is accepted at a confidence level $1-\beta$. 
Assuming a confidence level of $90\% $ the alternative hypothesis H$_1$ is excluded for $\beta =  0.1$. It is thus possible to find a value of signal with $1 - \beta = 0.9$, which leads to $Z_{90}\geq 3$. For this case, an experiment has a $90\% $ probability to detect at least a $3\sigma$ signal.

Concretely, we generate between $250$ and $500$ Monte-Carlo data sets for the signal plus background hypothesis $\mathrm{H}_1$ for each value of DM mass $m_{\chi}$ or progenitor mass $m_X$ and for each cross section $\sigma_n$ or lifetime $\tau_X$. These data sets determine the significance distribution. 
The procedure is repeated for a range of cross sections or lifetimes until we obtain $Z_{90}=3$ for $\beta = 0.1$ which corresponds to a $3 \sigma$ discovery potential at $90\%$~C.L.. We repeat this procedure until the entire parameter space is mapped out. The task at hand is now to define appropriate signal and background hypothesis which can be tested using the profile likelihood ratio formalism. 

\paragraph*{WIMP signal/no DR background:} In order to \textit{discover} a WIMP signal, the background only hypothesis ${\rm H}_0$: $\sigma_n=0$ must rejected; the alternative hypothesis is ${\rm H}_1$: $\sigma_n>0$. The first scenario to consider here is the complete absence of any DR. This allows us to recover the standard result on the neutrino floor in the  $(\sigma_n,m_\chi)$-plane~\cite{Billard:2013qya}. In other words we take~$\alpha = {\rm DM}, \nu_j$ in Eq.~(\ref{eq:Levents}) and  $\beta = \nu_j$ in Eq.~(\ref{eq:Lflux}) to minimze  the likelihood ratio~(\ref{eq:test_statistic}). 

\begin{figure*}[tb]
  \includegraphics[width=\columnwidth]{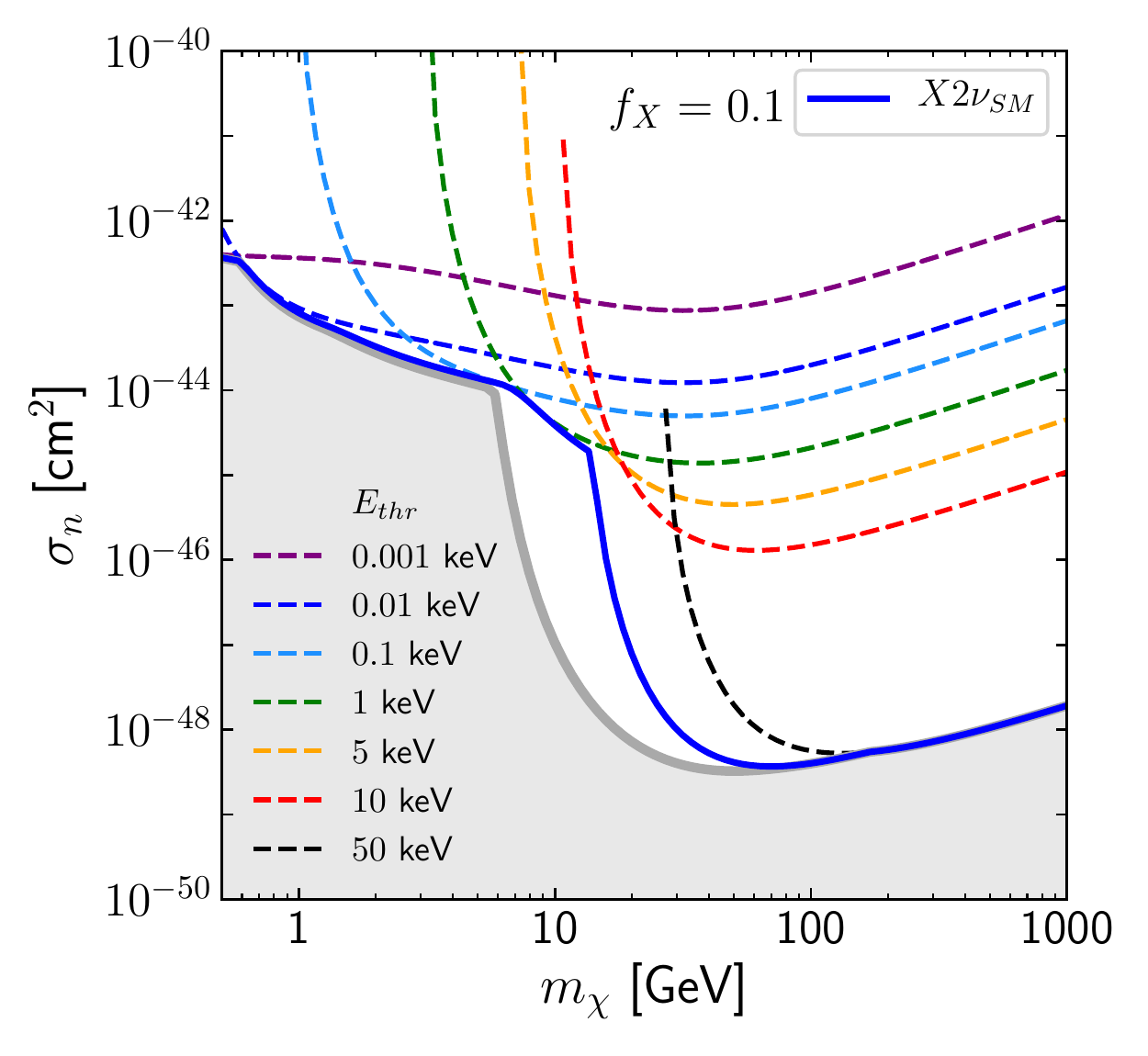}%
\includegraphics[width=\columnwidth]{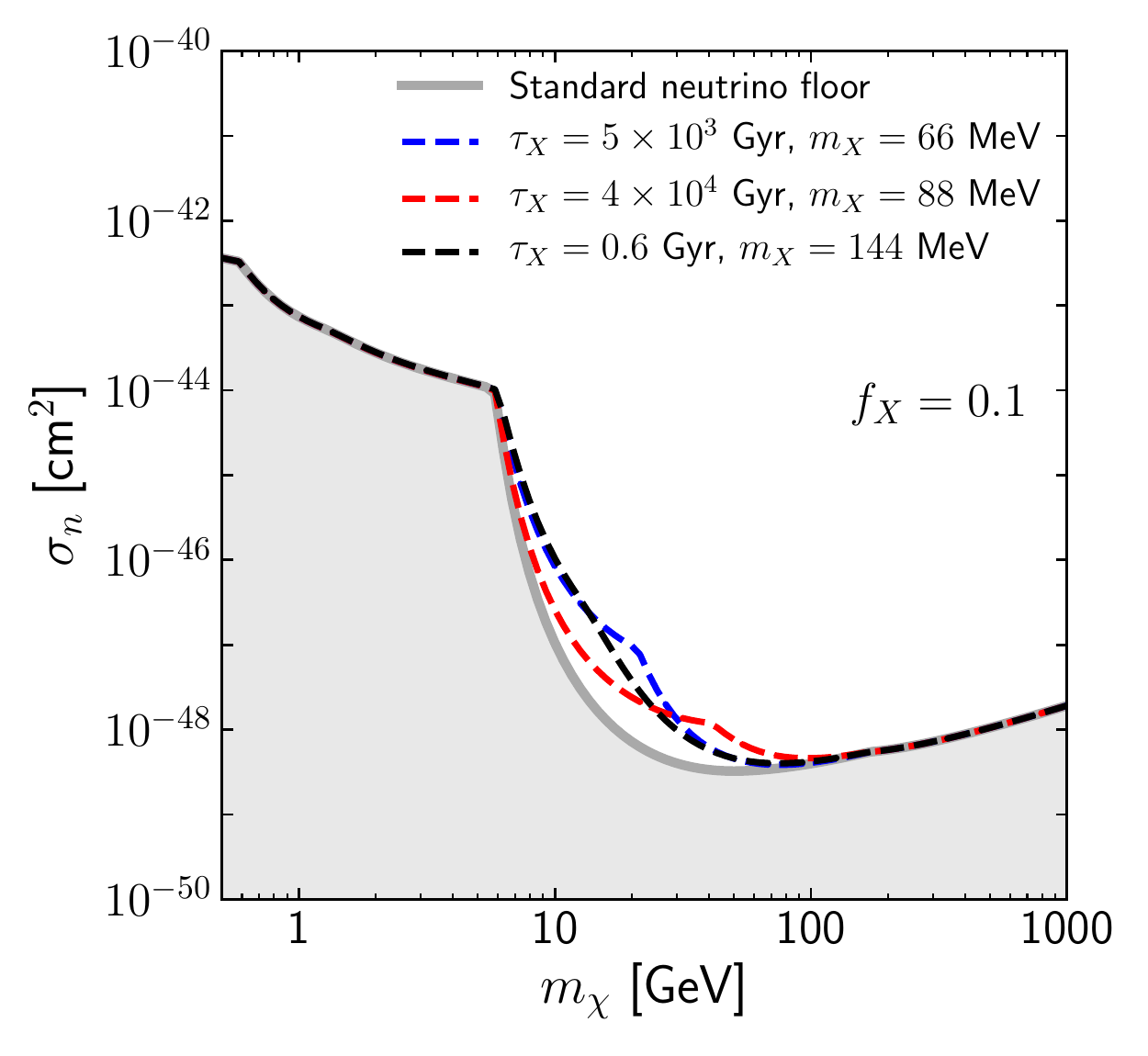}
  \caption{The DM exclusion potential in the presence of DR; the gray region shows the standard result without DR \textit{Left panel:} the various dashed lines show the limits for varying threshold energy for DR induced by a progenitor of $m_X=60\,\MeV$ and $\tau_X= 10$~Gyr. The thick blue curve as the minimum of all curves is the modified neutrino floor. \textit{Right panel:} dependence of the neutrino floor on progenitor mass and lifetime for exemplary combinations.}
\label{fig:excl_WIMP}
\end{figure*}

 \begin{figure}[tb] \includegraphics[width=0.97\columnwidth]{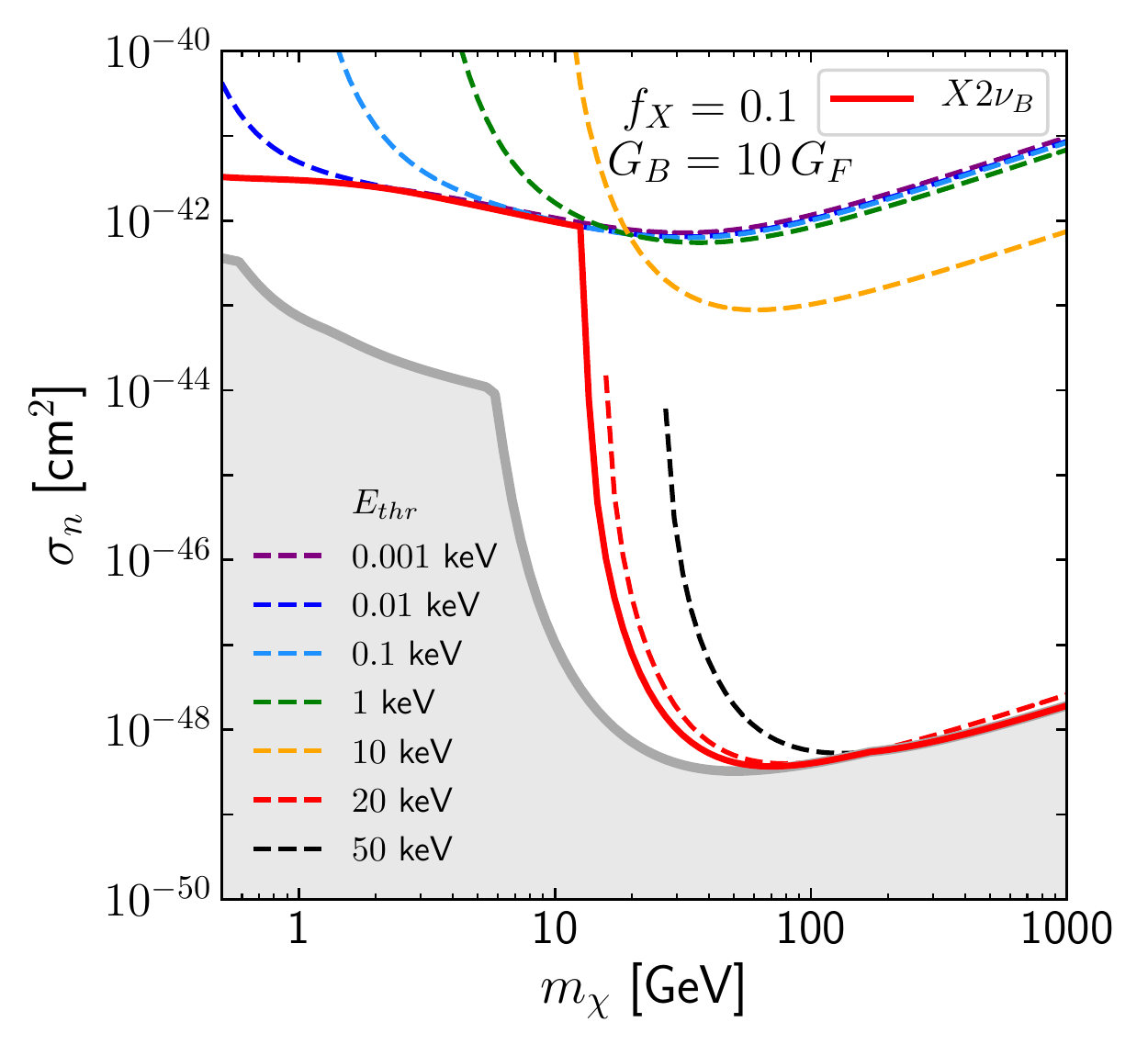}
   \caption{The DM exclusion potential in the presence of DR in form of baryonic neutrinos $\nu_B$ with $G_B=10G_F$; the gray region shows the standard result without DR. The various dashed lines show the limits for varying threshold energy for DR induced by a progenitor of $m_X=60\,\MeV$ and $\tau_X= 10$~Gyr. The red blue curve as the minimum of all curves is the modified neutrino floor.}
\label{fig:excl_WIMP_bar}
\end{figure}

\paragraph*{WIMP signal with DR background:} The next step is to switch on DR as an additional source of background and study the influence on the WIMP discovery region. The background only hypothesis remains ${\rm H}_0$: $\sigma_n=0$ and the alternative hypothesis is again ${\rm H}_1$: $\sigma_n>0$, but now we take the sums in Eqs.~(\ref{eq:Levents}) and~(\ref{eq:Lflux}) to include DR, \textit{i.e.}~$\alpha = {\rm DM},\, X2\nu,\, \nu_j$ and  $\beta = X2\nu,\, \nu_j$, respectively. We fix progenitor mass and lifetime for this procedure.

\paragraph*{DR signal with standard neutrino backgrounds:} The final alternative is to explore the DR discovery potential of   direct detection experiments. Hence, we treat the DR recoils as signal in the presence of standard neutrino backgrounds but in absence of DM-induced events. The null hypothesis is ${\rm H}_0$: $\sigma_n=0$ and $\tau_X \to \infty$ and the alternative hypothesis becomes ${\rm H}_1$: $\sigma_n=0$ and finite $\tau$.  We therefore take the sums in Eqs.~(\ref{eq:Levents}) and~(\ref{eq:Lflux}) with \textit{i.e.}~$\alpha = X2\nu,\, \nu_j$ and  $\beta =\nu_j$, respectively. In the baryonic neutrino scenario, we fix $G_B$ to an exemplary value.

The formalism described above is not specific to any particular detector. As was shown above, DR fluxes from DM decay can compete with standard solar neutrino fluxes only above the ${}^8$B endpoint energy. Dark radiation  introduces changes to the standard picture in the recoil energy region where the large WIMP detectors have ample sensitivity and ultra-low thresholds are of little benefit. In the following, we will hence explore the formalism on the basis of ton-scale xenon experiments like the ones mentioned in the introduction.

\section{Results}
\label{sec:results}
\subsection{Exclusion limits}

\begin{figure*}[tb]
  \includegraphics[width=\columnwidth]{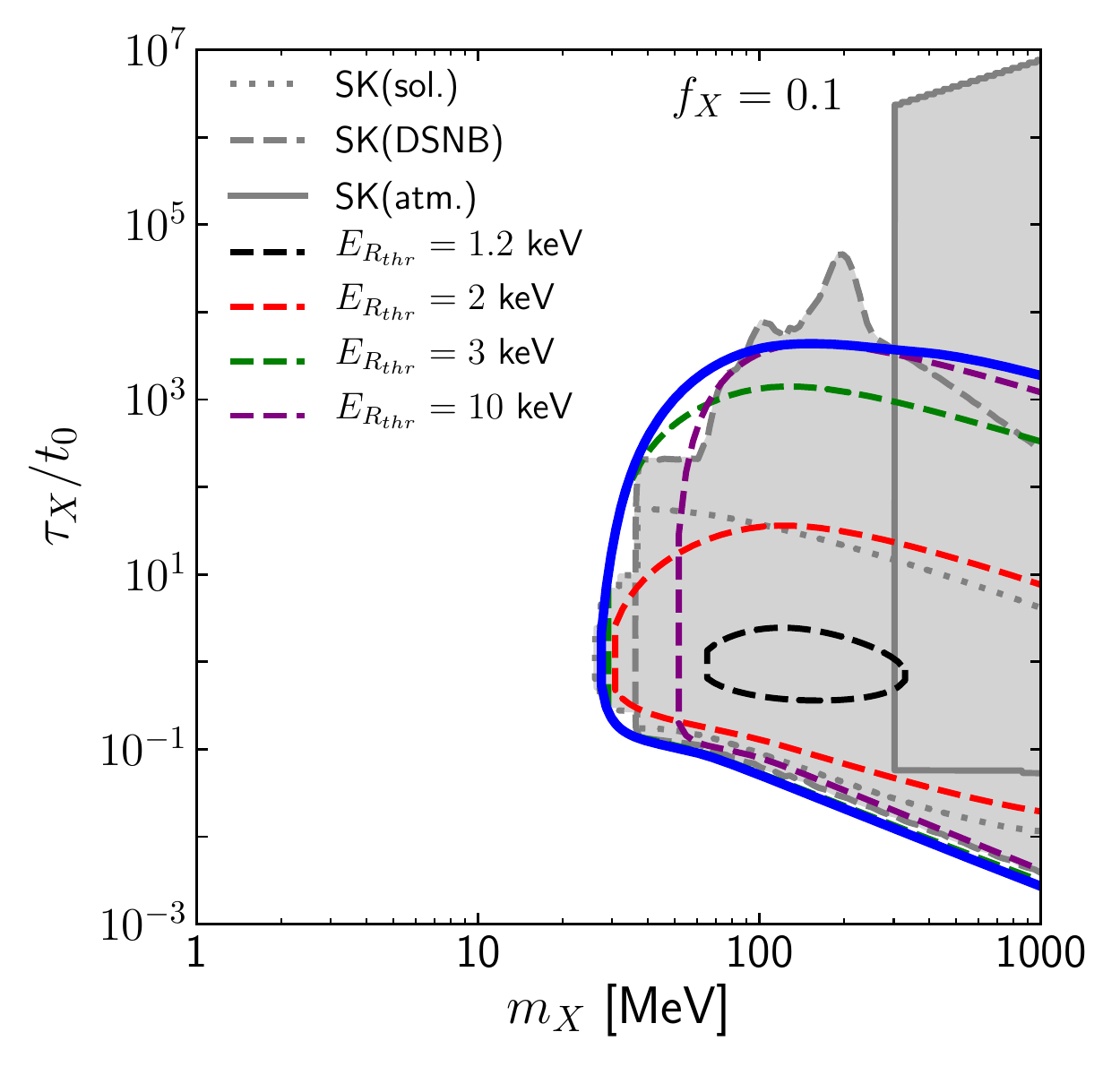} \includegraphics[width=\columnwidth]{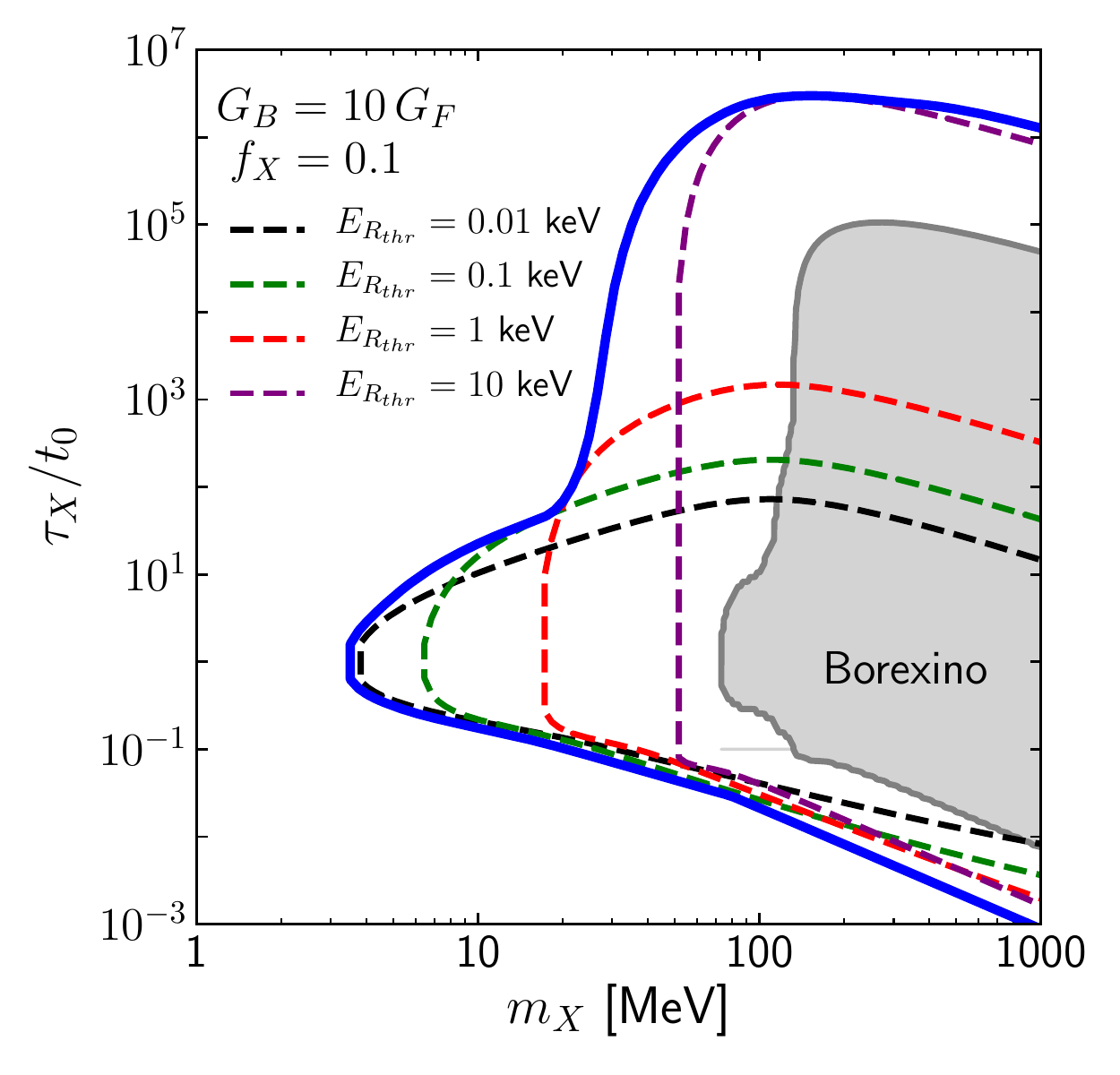}
  \caption{Obtainable exclusion limits, for DR in form of SM neutrinos (baryonic neutrinos) in the left (right) panel, assuming $f_X=0.1$. The various dashed curves assume different nuclear recoil energies as labeled. The thick blue line is the maximum of all curves. The gray areas in the  left (right) panel show excluded regions from SK (Borexino) and are taken from~\cite{Cui:2017ytb}.}
\label{fig:contour_g}
\end{figure*}

To begin with, we establish the exclusion potential for DM when DR neutrinos become a source of additional background. In the remainder of the paper we shall focus on the example of a liquid xenon detector. The operational procedure is then as follows: given the presence of anomalous neutrino flux, we first compute the best DM limit at a fixed threshold by finding the exposure for which  one background event as explained in Sec.~\ref{sec:stats_exclusion} is seen. In a second step, the threshold is varied and the minimum of all the exclusion curves, i.e.~the minimum value of $\sigma_n$ for each $m_\chi$, is obtained. The resulting contour in the $(\sigma_n,m_\chi)$ plane then represents the optimal DM exclusion potential of the experiment when irreducible backgrounds are not dealt with.

Figure~\ref{fig:excl_WIMP}, shows the results of the procedure. In the left panel, the results for a fixed progenitor mass of 60~MeV, and lifetime of 10~Gyr decaying into SM neutrinos are shown ($X2\nu_{SM}$ case). The various dashed curves show the best limit on $\sigma_n$  for various assumed thresholds, from $1$~eV to $50$~keV, as labeled. The blue solid line is the minimum of all curves and represents the optimal exclusion potential. The curves are to be compared with the gray line and shaded region which represents the standard exclusion potential. As can be seen, modifications to the standard result are obtained in the regions between $6-50$~GeV WIMP mass, which can be several orders of magnitude.
The right panel of Fig.~\ref{fig:excl_WIMP} shows the modified exclusion potential in the presence of SM DR neutrinos for currently allowed $(m_X, \tau_X)$ combinations: the parameter point in the left panel is already challenged from measurements of atmospheric fluxes and searches for DSNB neutrinos by Super-Kamiokande and the parameter points in the right panel give the largest, albeit modest modification to the exclusion potential (see below). 

The general trend in both panels of Fig.~\ref{fig:excl_WIMP} is that the exclusion potential  is shifted to the right off the standard ${}^8$B shoulder. How far that shift goes, depends on kinematics. For a progenitor mass of $m_X = 60\,\MeV$, DR neutrinos induce recoils that compete with DM only below $m_\chi \lesssim 50\,\GeV$. The exclusion potential for heavier DM is then only challenged by the standard, more energetic atmospheric fluxes. If we are to entertain the possibility of $m_X\gg 100\,\MeV$ into the GeV-regime, the floor would be lifted by several orders of magnitude in the electroweak scale DM mass regime.

 Figure~\ref{fig:excl_WIMP_bar} shows the modification to the standard exclusion limit (gray) from DR in form of baryonic neutrinos ($X2 \nu_B$ case). The red solid line is the optimal exclusion potential with an effective coupling $G_B=10\, G_F$. The sizable new coupling boosts the DR-induced nuclear recoil rate above the $^8$B neutrino recoil rate, weakening the minimal excludable DM-nucleon cross section by few orders of magnitude once $m_\chi < 50$~GeV.

The potential modifications entertained in Fig.~\ref{fig:excl_WIMP} and Fig.~\ref{fig:excl_WIMP_bar} have to be put in perspective with the current sensitivity of multi-ton scale neutrino experiments. This has been explored previously in Ref.~\cite{Cui:2017ytb} and the gray shaded regions in Fig.~\ref{fig:contour_g} show the excluded regions in the DR progenitor plane $(m_X, \tau_X)$ from Super-Kamiokande and Borexino. The constraints included in the left panel of the Fig.~\ref{fig:contour_g} were derived from  measurements of solar- and atmospheric fluxes~\cite{Richard:2015aua,Abe:2010hy}, and from searches of DSNB neutrinos~\cite{Bays:2011si}. In the right panel of Fig.~\ref{fig:contour_g} the constraint was obtained by re-purposing a Borexino solar-axion search~\cite{Bellini:2012kz} to $\nu_B$-induced elastic proton recoils; for further details on both cases see~\cite{Cui:2017ytb}.
 The inside of the curves labeled according to the assumed threshold correspond to a combination of parameters where we find that DR sourced by a DM progenitor is discoverable in a direct detection experiment.  As before, we assume a 10\% fraction of decaying DM of type~$X$, $f_X=0.1$. 
 
 Also shown in Fig.~\ref{fig:contour_g} are the exclusion potentials for DR in form of SM neutrinos (left) and baryonic neutrinos (right) for threshold energies ranging from 0.01~keV up to 50~keV. In the region $\tau_X> t_0$ they are determined by the galactic and extra-galactic components whereas in the region $\tau_X< t_0$ only the extra-galactic flux contributes. The prospective regions are hence centered around $t_0$: for $\tau_X\gg t_0$ the flux is too small and for $\tau_X\ll t_0$ the extra-galactic flux becomes too soft. It is interesting to note that in the right panel of Fig.~\ref{fig:contour_g} below the {\it minimum} assumed threshold of 1.2~keV, DR in from of SM neutrinos cannot be excluded, as their induced events are superseded by solar neutrinos, especially the ones originating from the $^8$B reaction. On the flip side, as the threshold increases, the experiment becomes more sensitive to higher energy recoils generated by larger $m_X$. Hence, in contrast to the WIMP case, and because of the general limitations in the DR flux in form of SM neutrinos, ever lower thresholds as they are sought in many direct detection experiments are not a beneficiary factor.  In summary, as can be seen, except for relatively small regions located in the quadrant $m_X\lesssim 100\,\MeV$ and $\tau_X > t_0$, neutrino experiments carry stronger current constraining power to SM DR than DM direct detection experiments will every have.
 On the other hand, when we are to entertain new physics in the DR itself (in form of $\nu_B$), direct detection sensitive to lower masses $m_X \lesssim 20\ \GeV$ is gained as the threshold drops below~1~keV. But even with higher threshold values, and expanded region in $\tau_X$ at a given $m_X$ can be covered.

 \begin{figure*}[tb]
  \includegraphics[width=\columnwidth]{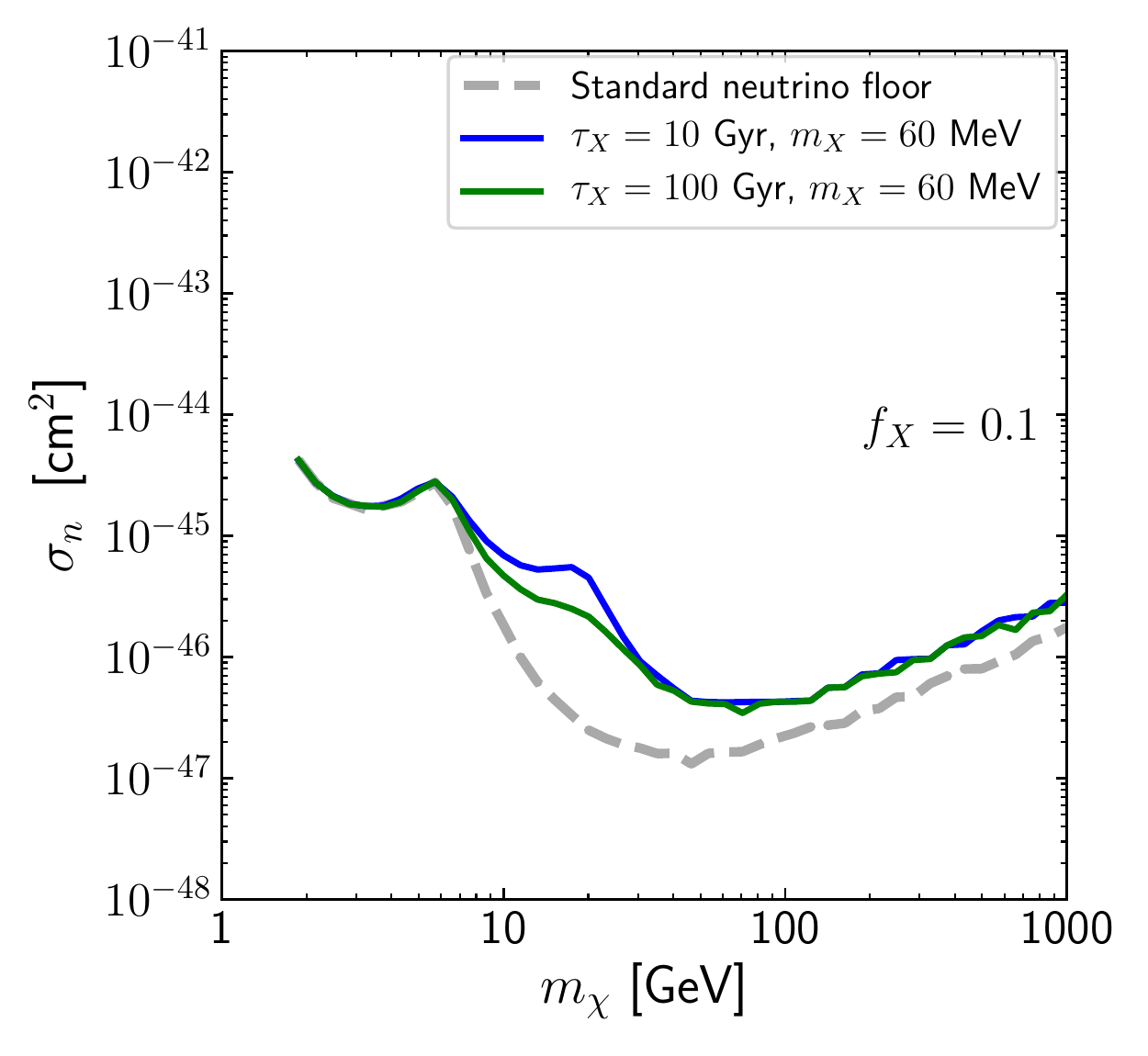}\hfill
  \includegraphics[width=\columnwidth]{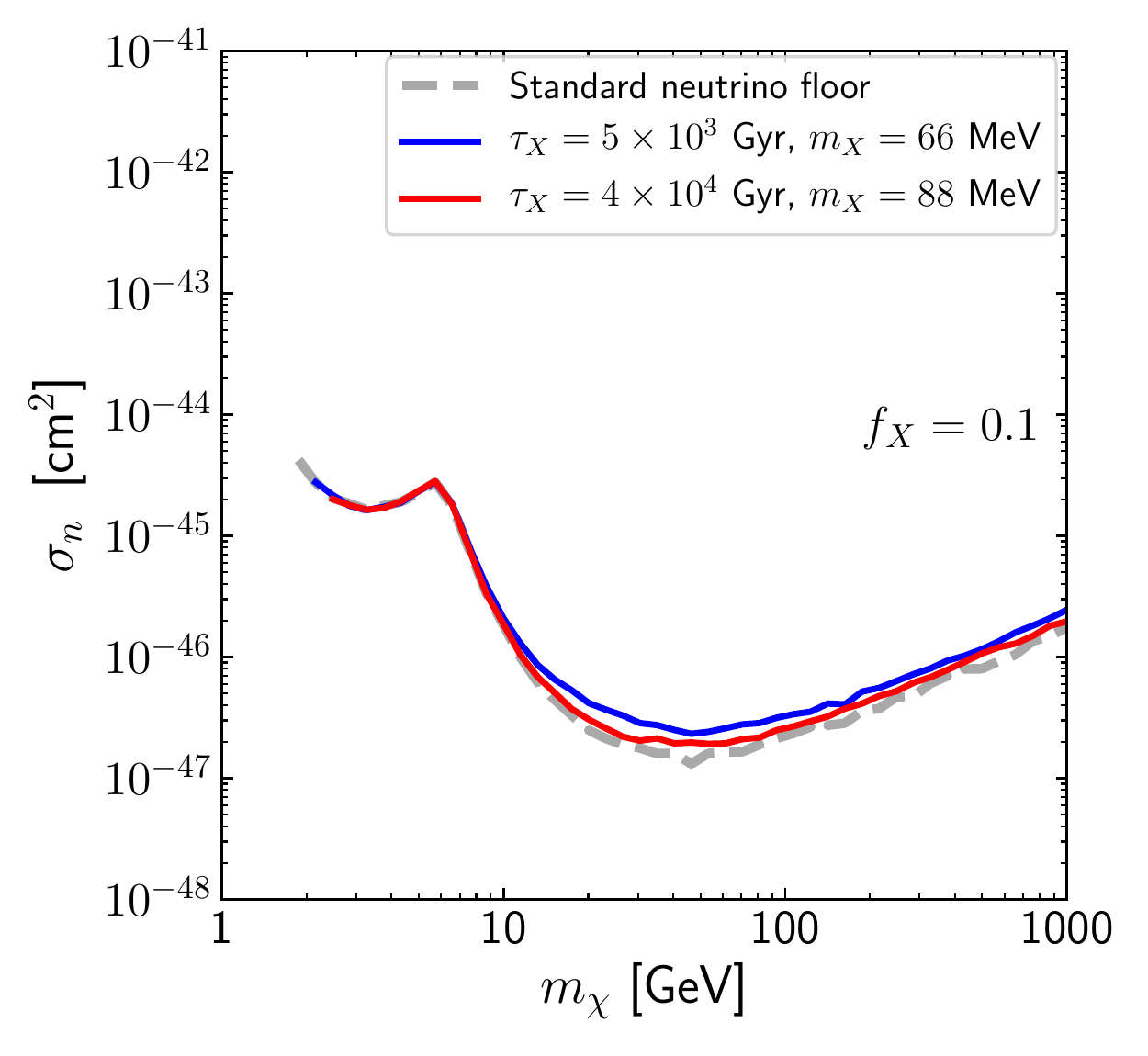}
\caption{The modified neutrino floor at $90\%$ C.L.~in case of $X2\nu_{SM}$ for a threshold $E_{R_\mathrm{thr}}=0.1$~keV and exposure of 1~ton-yr \textit{Left panel:} for progenitor $X$ mass of 60 MeV and lifetime of $10$~Gyr \textit{Right panel:} for exemplary allowed combinations $(\tau_X,_X)$. For reference, we also show the `standard' neutrino floor for the DM-only signal hypothesis.}
\label{fig:discovery_WIMP}
\end{figure*}

\begin{figure}[tb] \includegraphics[width=0.97\columnwidth]{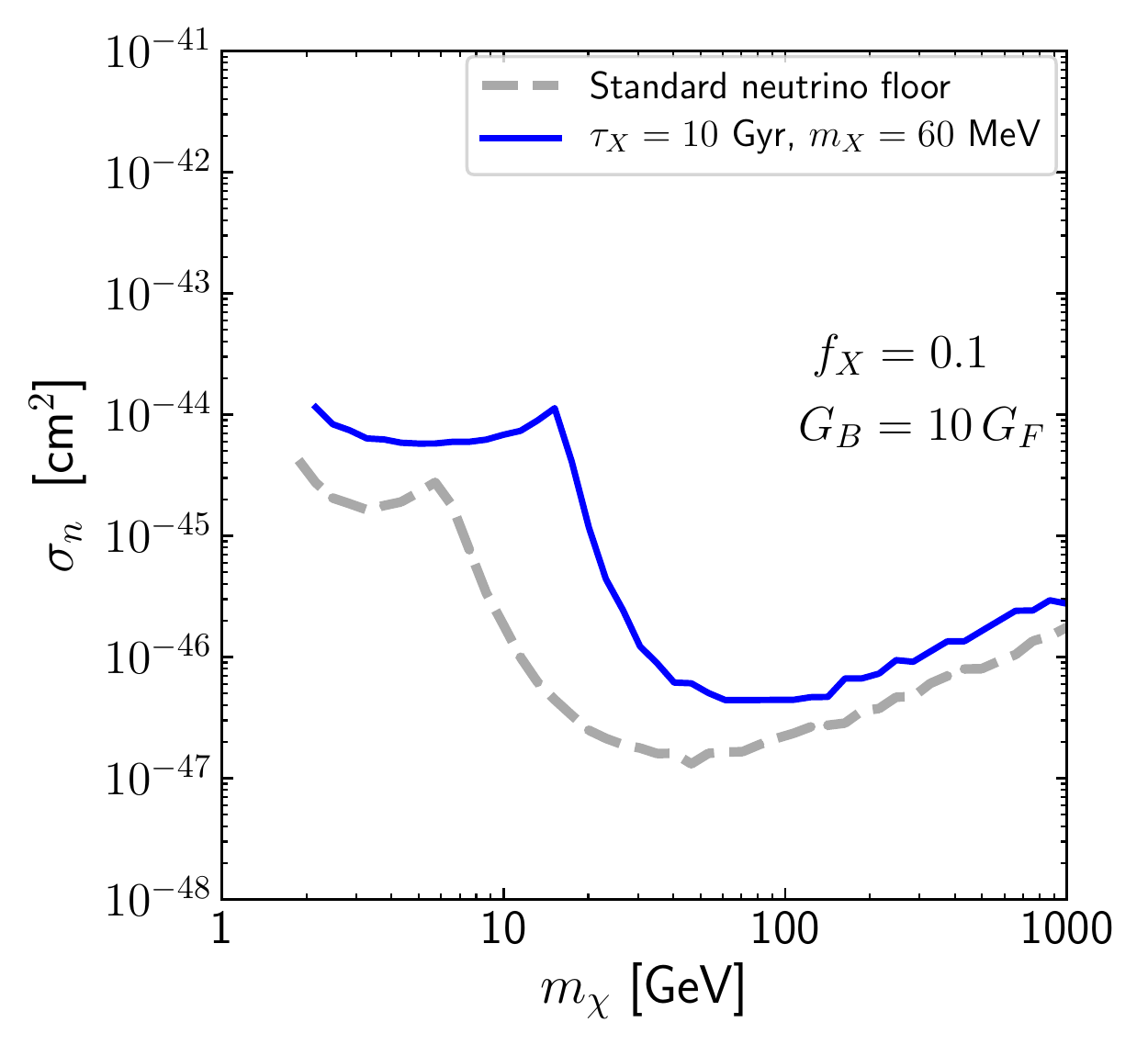}
   \caption{The modified neutrino floor at $90\%$ C.L.~in case of $X2\nu_{B}$ for a threshold $E_{R_\mathrm{thr}}=0.1$~keV and exposure of 1~ton-yr for progenitor $X$ mass of $60$~MeV and lifetime of $10$~Gyr.
}
\label{fig:discovery_WIMP_bar}
\end{figure}

\begin{figure*}[tp]
  \includegraphics[width=\columnwidth]{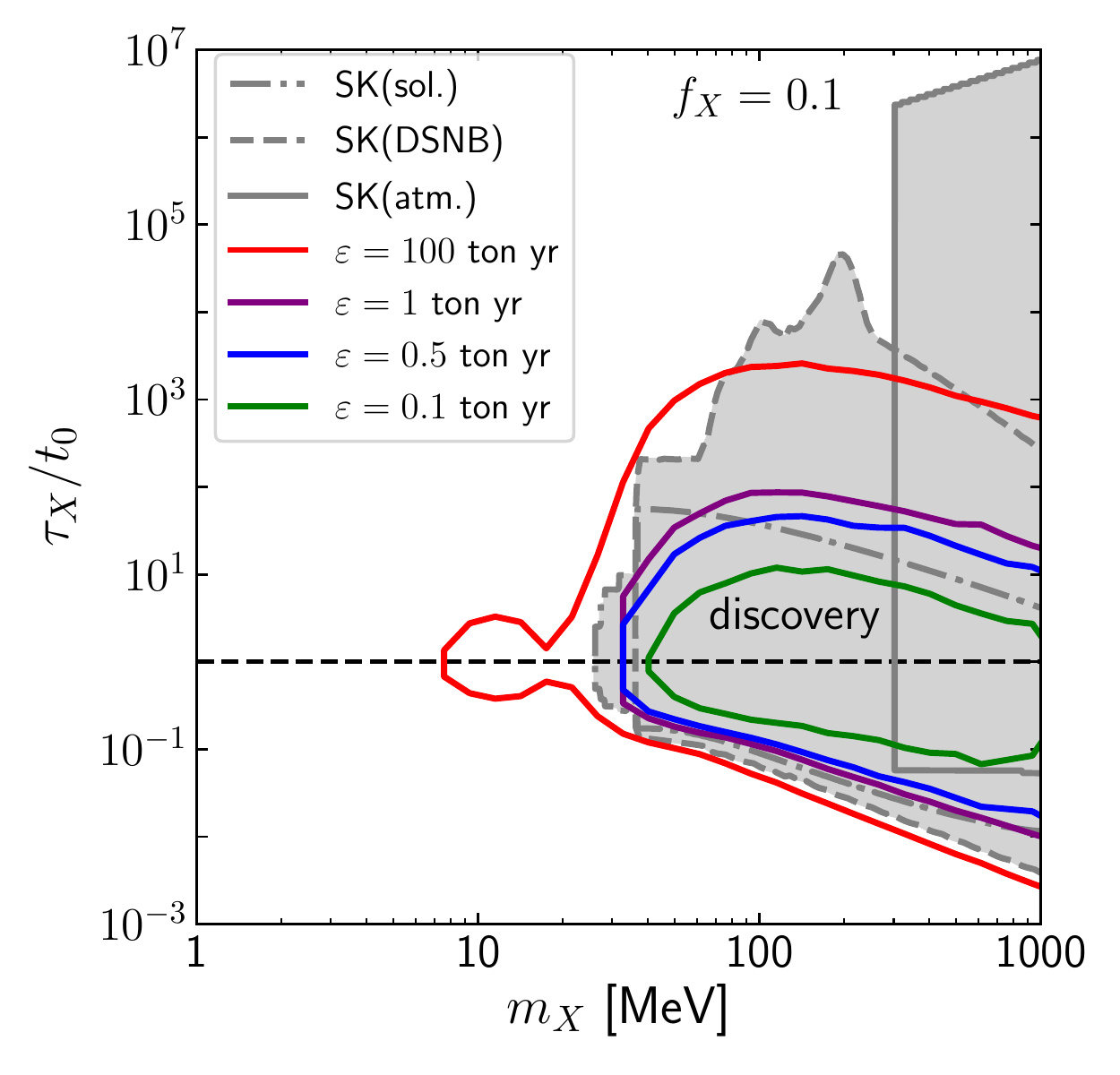}%
  \includegraphics[width=\columnwidth]{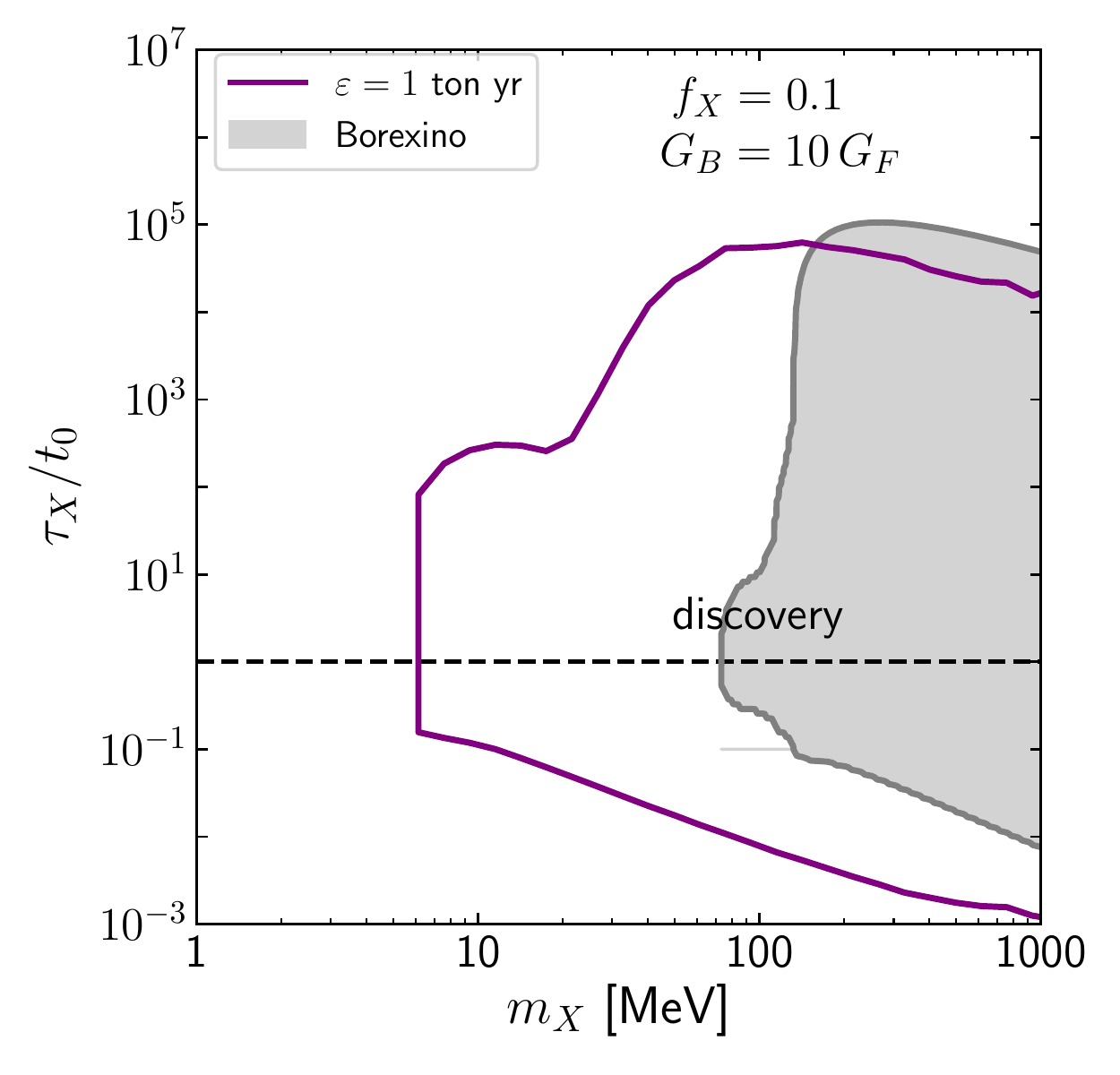}%
\caption{Discovery limits at a $90\%$ C.L. for a threshold $E_{R_\mathrm{thr}}=0.1$~keV \textit{Left panel:} for DR in form of SM neutrinos \textit{Right panel:} for DR in form of baryonic neutrinos.}
\label{fig:discovery_contour_g}
\end{figure*}

\subsection{Discovery potential}

We now turn our attention to the prospects of making a discovery of either DM or DR in the presence of irreducible backgrounds which are to be faced in the coming generation of direct detection experiments. Charting out the discovery regions will be achieved using the likelihood approach outlined in Sec.~\ref{sec:stats_discovery}.  Similarly to the previous section, we attempt at discovering WIMP recoils in presence of standard plus DR neutrino sources and we also estimate the discovery potential for DR itself in presence of standard neutrino backgrounds. Importantly, such exercise requires specification of an experiment's exposure and threshold and is hence specific to these assumptions. In the following we choose $\varepsilon = 1$~ton-yr and $E_{\rm thr} = 0.1$~keV to normalize on current collected data sets combined with an aggressive assumption on nuclear recoil threshold.

In the left panel of Fig.~\ref{fig:discovery_WIMP}, we show the DM discovery potential for $X2\nu_{SM}$ for a progenitor mass of 60~MeV and for progenitor lifetimes of 10~Gyr (blue line) and 100~Gyr (green line).  As in the previous sections, modifications to the standard result (dashed line) appear for $m_\chi  \gtrsim 6~\GeV$, when DR fluxes come to dominate the event rates. The modification of discovery limits are seen up to DM masses of 1~TeV. 
In the right panel of Fig.~\ref{fig:discovery_WIMP} we show the discovery limits for $X2\nu_{SM}$ for the same allowed combinations $(\tau_X,m_X)$ as chosen in Fig.~\ref{fig:excl_WIMP} (right). As can be seen, the maximal allowed DR signal is at most comparable to the SM neutrino backgrounds and only mildly alters the standard neutrino floor (gray dashed). 

Figure~\ref{fig:discovery_WIMP_bar} shows the modified DM discovery potential for $X2\nu_{B}$ for a progenitor mass of $60$~MeV, for a lifetime of 10~Gyr and $G_B = 10 G_F$. For this case, we switch to a binned likelihood approach~\eqref{eq:LeventsBin} with $N_{\rm bin}=50$. As already shown in Fig.~\ref{fig:excl_WIMP_bar}, the same behaviour of the curve can be seen. Compared to Fig.~\ref{fig:discovery_WIMP}, owing to the increased strength of interaction, the neutrino floor is also raised in the low mass region left of the $^8$B shoulder ($m_\chi \lesssim 6\,\GeV$). 
In both Figs.~\ref{fig:discovery_WIMP} and \ref{fig:discovery_WIMP_bar}, the modification 
 of discovery limits are seen up to the largest shown DM masses of 1~TeV. This is in contrast to the exclusion potentials of the previous section. For the latter, only the number of events for varying threshold (up to 50~keV) enter, but not their detailed spectral shape. 
 In this section, we fix the threshold to $E_{\rm thr}=0.1$~keV so that DR spills into the standard induced events at any rate, also affecting the discovery potential for $m_\chi\gtrsim 50,\GeV$.

Finally, we also present the prospects of {\it discovering} DR at direct detection experiments in Fig.~\ref{fig:discovery_contour_g}. We  consider exposures up to 100~ton-yr for a threshold of 0.1~keV for DR in form of SM neutrinos (left panel) and baryonic neutrinos (right panel). For the extreme case of 100~ton-yr exposure, reflective of the sensitivity of future multi-ton xenon detectors such as DARWIN~\cite{Aalbers:2016jon}, we switch to a binned likelihood approach~\eqref{eq:LeventsBin} with $N_{\rm bin}=50$.  The limiting factors for the reach can be understood by analysing the most dominant background fluxes. In the bulk of the $(m_X,\tau_X)$ parameter space, hep neutrinos constitute the dominant background to~DR. This is because hep neutrinos generate a large range of recoil energies with appreciable flux. Around $20$~MeV progenitor mass, the $^8$B neutrinos become the limiting factor. 
As can be seen in the left panel of Fig.~\ref{fig:discovery_contour_g}, it is not possible to make a $3\sigma$ discovery of DR in form of SM neutrinos, unless a 100~ton-yr exposure is assumed (and paralleling progress with neutrino detectors is neglected.)
When $\nu_B$ are invoked (right panel), they can still be discovered in a wide mass range with only 1~ton-yr exposure.

\section{Conclusions}
\label{sec:conclusions}

In the foreseeable future, DM direct detection experiments will face irreducible background in the form of neutrino-nucleus elastic scattering.  The responsible neutrino fluxes originate from the sun and to a smaller degree from the atmosphere or from the cosmological SN history.  In this work, we first explore the impact on the DM discovery and exclusion potential at next generation direct detection searches in the presence of new, anomalous neutrino fluxes. The latter are sourced by a per-cent fraction of DM that may decay with arbitrary lifetime after recombination. We consider either SM or baryonic neutrinos $\nu_B$ as DR. The latter designate a semi-sterile fourth species that interacts through gauged baryon number. In a next step, we also address the question on the detectability of DR itself.

The DR fluxes principally fall into two categories, either constituting neutrinos that travel cosmological distances or neutrinos that originate from DM decay in the galaxy. Whereas the latter requires $\tau_X\gtrsim t_0$ as otherwise the DR-generating sub-component of DM would have already decayed, the former contributes for any $\tau_X$ and we provide a new closed-form expression for its total flux. In the detector, the neutrinos then interact coherently with the nucleus, with a cross section that approximately scales as $(A-Z)^2$ and $A^2$ for SM and baryonic neutrinos, respectively. The benefit of DR in form of SM neutrinos is that its interactions are known. In turn, DR in form of $\nu_B$ allows for effective interactions that can be stronger than in the SM, $G_B>G_F$, boosting the principal importance of the DR signal.

We then establish the DM exclusion potential of future  WIMP searches. We do so on the example of a ton-scale liquid xenon detector, varying its threshold and assuming an exposure such that the first neutrino events will be seen. In principle, modifications to the standard result, i.e.~without DR, can be several orders of magnitude. For DR in form of SM neutrinos, we find that the best sensitivity to the DM-nucleon cross section is at most weakened by about an order of magnitude once the scenario is subjected to existing constraints from SK data. We find that the limitations to exclude values of $\sigma_n$ originating from the solar $^8$B flux remain unaltered. This is owed to the principal cap on the size of the DR flux. In turn, for $m_\chi\gtrsim 6\,\GeV$, the region in the DM parameter space that is subject to background from atmospheric, hep, and DSNB neutrinos is affected substantially, reaching its largest modification for $m_\chi\simeq 20-30\,\GeV$. This is owed to the fact that neutrinos of $30\,\MeV$ have similar recoil characteristics as DM of that mass range, while at the same time the allowance on a non-standard DR flux is largest (DSNB search window). The region $m_\chi\gtrsim 50\,\GeV$ becomes again unaltered, as any modification to it would require a progenitor mass $m_X\gg 100\,\MeV$; the DR then becomes amply visible in the standard neutrino detectors, excluding  fluxes in excess of the standard ones. Finally, considering $\nu_B$ as DR with $G_B>G_F$ we find that the entire region $m_\chi \lesssim 50,\GeV$ can be strongly affected, and the DM exclusion potential is modified by many orders of magnitude. These general conclusions carry over from exemplary parameter points to the $(m_X,\tau_X)$ plane, shown in Fig.~\ref{fig:contour_g}.

In a second part, we then study the discovery potential for DM in the additional presence of DR and for DR in the presence of SM neutrino backgrounds. For this we embark into the profile likelihood method which allows for a concise test of the various hypotheses.
For various combinations of detection threshold and exposure, we build a statistical sample by Monte-Carlo generating a large set of mock recoil spectra that are subsequently used in the maximization of the likelihood function. Through this numerically expensive procedure we are able to obtain a DR-modified ``neutrino floor'', i.e.~the boundary in the $(m_\chi,\sigma_n)$ parameter space for the ability to detect DM with $3\sigma$~significance when neutrino backgrounds cannot be rejected. We find that DR in form of SM neutrinos only marginally affect  our prospect for DM discovery (by less than a factor of two), once complementary constraints on the progenitor parameter space are taken into account. For the $\nu_B$ case, we find that the discovery potential for DM can be significantly weakened. 

In a final step, we then ask the question on the detectability of DR itself. Here, standard neutrino sources are the irreducible background, and we find that SM DR cannot be discovered in direct detection experiments unless a futuristic exposure of 100~ton-yr is assumed. The coverage in new parameter space is modest (and neglects the paralleling advances in neutrino detectors). In contrast, direct detection experiments bear better potential to discover non-standard neutrino interactions of DR, as exemplified in the $\nu_B$ case, see Fig.~\ref{fig:discovery_contour_g}. Here progenitor masses down to $m_X = 10\,\MeV$ and lifetimes as large as $10^5\, t_0$ can be discovered.

With the advance of multi-ton scale direct detection experiments, we are entering the ``end-game'' of WIMP detection. Neutrinos from standard sources will become an irreducible background and ultimately limit our ability to push for ever smaller event rates and better DM sensitivity. It is hence only timely to ask, what other irreducible background there could be. Here we investigated the perfect possibility that our Universe is filled with MeV-scale DR that traces back to the instability of (a component of) DM. 
We find that standard expectations can be altered, and---in the presence of new particles and interactions---in a significant manner. In the latter case, DM direct detection experiments may then be turned into DR detectors instead.

\paragraph*{Acknowledgments}
We thank Y.~Perez-Gonzalez, and W.~Waltenberger for useful discussions. MN is supported by the FWF Research Group grant FG1. JP is supported through the New Frontiers Program by the Austrian Academy of Sciences. SK is supported by Elise-Richter grant project number V592-N27. 

\appendix
\begin{widetext}
 
\section{Analytic form of \boldmath${dR_{X2\nu}}/dE_R$}
\label{app:ana_diff_rate}

For 2-body decays into massless neutrino DR, the energy-differential recoil rate given in Eq.~(\ref{eq:SM_nu_recoil}) can be integrated analytically. For the extragalactic signal with the flux given in~\eqref{eq:egal_flux} we split the integral over neutrino energy into the two terms that comprise the cross section~\eqref{eq:nu_nucleus_sigma},
\begin{equation}
\begin{aligned}
\frac{dR^\mathrm{\, e.gal}_{X2\nu}}{dE_R}  & = 
N_T \frac{Q_W^2G_F^2 m_N F(q)^2}{ 4 \pi } \left( \int_{E_\mathrm{min}}^{E_\mathrm{in}} d E_\nu \; \frac{d \Phi_{\nu, \mathrm{e.g.}}}{dE_\nu} -\frac{ E_R m_N}{2} \int_{E_\mathrm{min}}^{E_\mathrm{in}} d E_\nu \; \frac{d \Phi_{\nu, \mathrm{e.g.}}}{dE_\nu} \frac{1}{E_\nu^2} \right),
\end{aligned}
\end{equation}
where $E_\mathrm{in}=\frac{m_X}{2}$ is the injection energy and $E_\mathrm{min}= \sqrt{\frac{E_R m_N}{2}}$ is the minimum energy to produce a recoil $E_R$. The  integrals are solved by substitution, \begin{equation}
    u=\sqrt{\left(\frac{E_\mathrm{in}}{E_\mathrm{min}}\right)^3 \frac{\Omega_M}{\Omega_\Lambda} +1}, \quad w=\frac{u+1}{u-1},
\end{equation}
and we obtain the expression,
\begin{equation}
\begin{aligned}
\label{eganalytic}
\frac{dR^\mathrm{\, e.gal}_{X2\nu}}{dE_R} & =N_T \frac{Q_W^2G_F^2 m_N F(\left|\vec{q}\,\right|)^2}{ 4 \pi } N_\nu  \frac{f_X \Omega_\mathrm{dm} \rho_\mathrm{crit}}{m_X } %
\left( x_\mathrm{min}^d- x_{\max}^d   
- 3d \left(\frac{\sqrt{2}\, \Omega_\Lambda}{\Omega_M} \right)^{\frac{2}{3}} \frac{E_R m_N}{E_\mathrm{in}^2} \frac{_2 F_1 \left( -\frac{1}{3}, \frac{1}{3}-d,\frac{2}{3};1-x \right)}{\sqrt[3]{-1+x }}  \Bigg|_{x_\mathrm{min}}^{x_{\max}}  \right).
\end{aligned}
\end{equation}
where %
\begin{equation}
d= -\frac{1}{3 H_0 \sqrt{\Omega_\Lambda} \tau_X}, \quad %
x_\mathrm{min}=\frac{\sqrt{\frac{E_\mathrm{in}^3}{E_\mathrm{min}^3}\frac{\Omega_M}{\Omega_\Lambda} +1}+1}{\sqrt{\frac{E_\mathrm{in}^3}{E_\mathrm{min}^3}\frac{\Omega_M}{\Omega_\Lambda} +1}-1} , \quad x_{\max}= \frac{\sqrt{\frac{\Omega_M}{\Omega_\Lambda} +1}+1}{\sqrt{\frac{\Omega_M}{\Omega_\Lambda} +1}-1}.
\end{equation} 
The total flux $\Phi_{X2\nu}$ in Eq.~\ref{eq:PhiEg} is obtained from the first two terms in the bracket of~\eqref{eganalytic} by setting $x_{\min}=1$. 
For the galactic component the integral with the flux given in~\ref{eq:gal_flux} can be evaluated directly,
\begin{equation}
\begin{aligned}
\label{eq:dR_gal}
\frac{dR^{\,\textnormal{gal.}}_{X2\nu}}{dE_R}  & =N_T  \frac{Q_W^2 G_F^2 m_N F(\left|\vec{q}\,\right|)^2}{ 4 \pi } N_\nu   \frac{f_X \, e^{-\frac{t_0}{\tau_X}}}{\tau_X m_X } r_\odot\rho_\odot \langle{J_\mathrm{dec}(\theta)}\rangle \left(1- \frac{2 E_Rm_N}{m_X^2} \right).
\end{aligned}
\end{equation}
The DR rate is then the sum of both, galactic and extragalactic contributions. The baryonic neutrino rate is calculated in the same way replacing ${Q_W^2 G_F^2}/4 \pi$ by ${A^2G_B^2}/{2 \pi}$.
\end{widetext}

\section{Details on discovery and exclusion potentials}
\label{sec:deta-disc-excl}

 \begin{figure*}[tb]
 \includegraphics[width=\columnwidth]{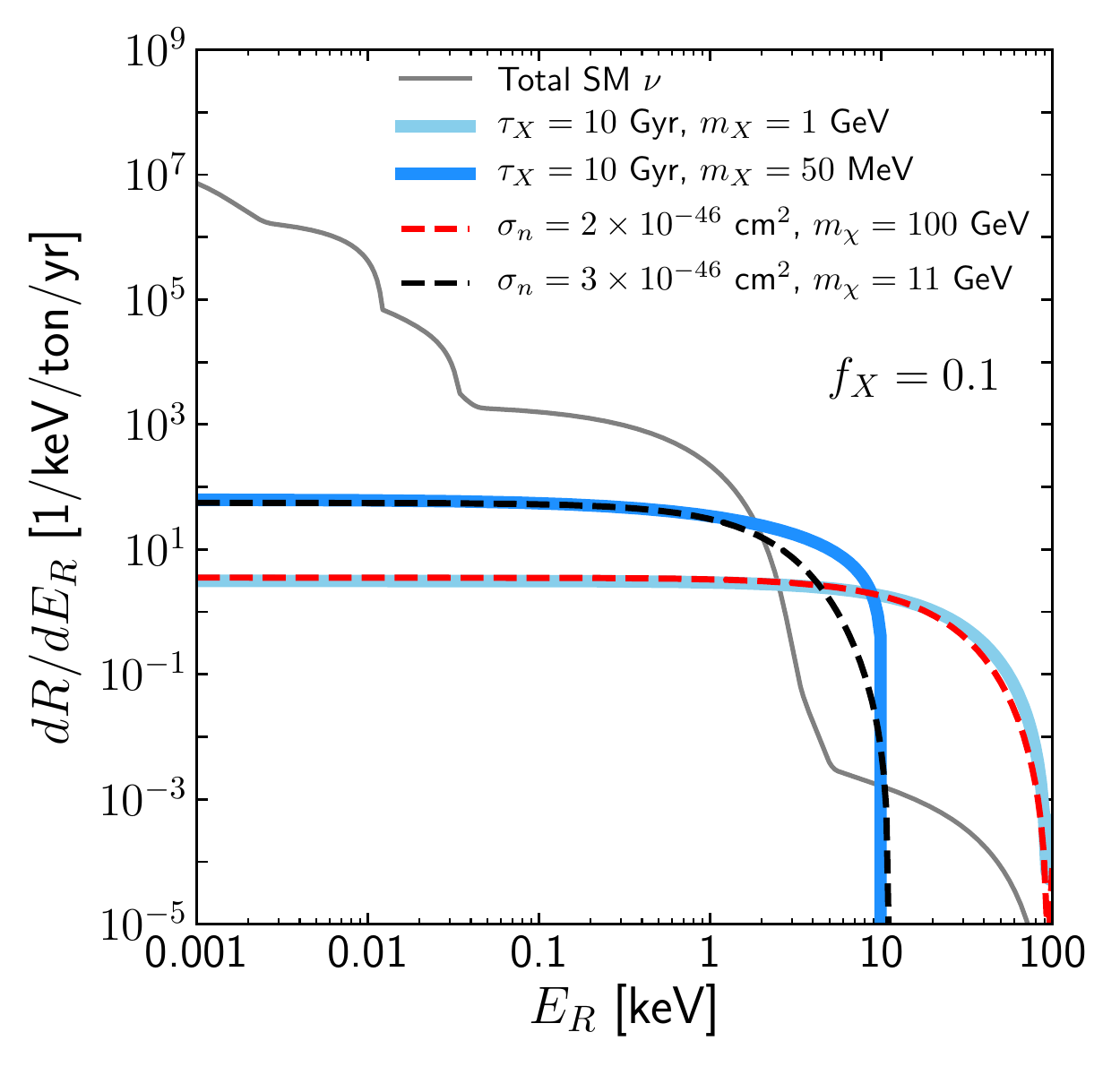}  \includegraphics[width=\columnwidth]{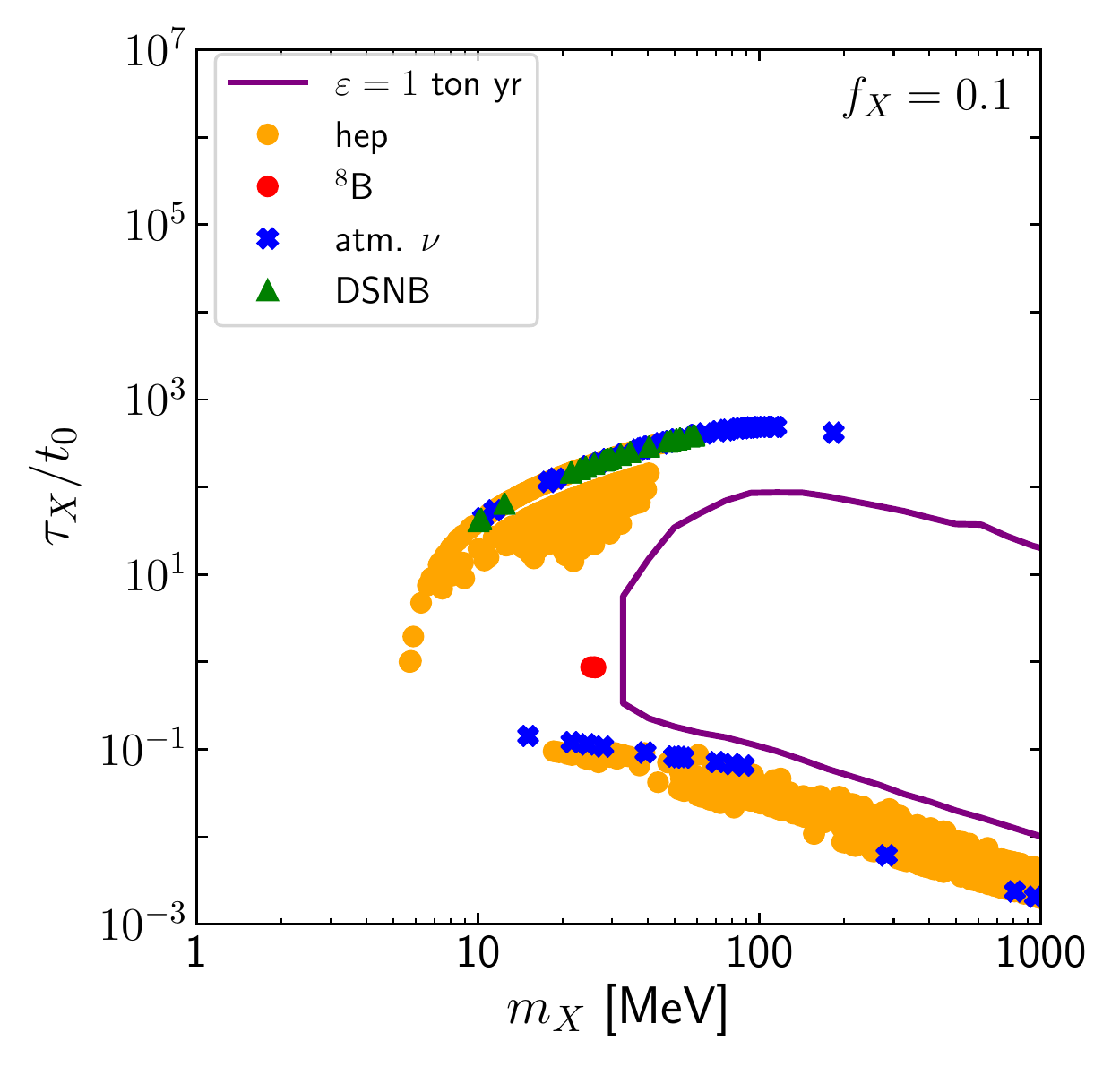}%
\caption{\textit{Left:} Best fit WIMP recoil spectra (dashed lines) which mimic the DR recoils (cyan solid lined) with progenitor lifetime of $10$ Gyr and progenitor mass of $50$ MeV (left) and $1$ GeV. The total standard neutrino recoil rates are also overlaid (solid grey). \textit{Right:}
Contributions from each neutrino source (best fits) shown by the points as labeled and determining the shape of the discovery limit for threshold $E_{R_\mathrm{thr}}=0.1$~keV and exposure of $\epsilon=1~$ton-yr.}
\label{fig:best_fit_WIMP}
\end{figure*}

Here we show some examples of exclusion and discovery limits, where SM neutrinos may mimic potential DR and DM signals in direct detection experiments. The reach of an experiment for a signal depends on the shape of the recoil rate and the number of events in relation to backgrounds. In this appendix we study cases where DR neutrinos exhibit some degree of similarity with DM-induced events.

In the left panel of Fig.~\ref{fig:best_fit_WIMP} we show two examples where  DM and DR  yield similar differential recoil spectra for two different progenitor masses 50~MeV and 1~GeV and DM masses as labeled such that the recoil endpoint energies match. For the lighter progenitor---and which is most prospective with respect to existing constraints (see above)---the DM and DR spectra are well distinguishable, with DR-induced events exhibiting a harder spectrum. For GeV-scale progenitors, the spectrum is cut off by the nuclear form factor $|F(q)|^2$ explaining thee similarity in the rate with an electroweak-scale mass DM particle. We note in passing, that in the latter case, additional inelastic nuclear channels are possible as the CM energy is in the tens of MeV regime.

In the right panel of Fig.~\ref{fig:best_fit_WIMP} we follow our usual procedure and simulate several thousand Monte-Carlo experiments for each neutrino source $\nu_j$ and calculate the distribution of best parameters $(m_X,\tau_X)$ by minimizing the likelihood from Eq.~(\ref{eq:Levents}) for $\alpha= X2\nu$. The discovery limit to SM DR neutrinos for an exposure of 1~ton-yr and threshold $E_{R_\mathrm{thr}}=0.1$~keV is shown. In addition, the dots as labeled are indicative of the standard neutrino fluxes that limit a further extended reach.

\bibliography{Refs}

\begin{thebibliography}{71}%
\makeatletter
\providecommand \@ifxundefined [1]{%
 \@ifx{#1\undefined}
}%
\providecommand \@ifnum [1]{%
 \ifnum #1\expandafter \@firstoftwo
 \else \expandafter \@secondoftwo
 \fi
}%
\providecommand \@ifx [1]{%
 \ifx #1\expandafter \@firstoftwo
 \else \expandafter \@secondoftwo
 \fi
}%
\providecommand \natexlab [1]{#1}%
\providecommand \enquote  [1]{``#1''}%
\providecommand \bibnamefont  [1]{#1}%
\providecommand \bibfnamefont [1]{#1}%
\providecommand \citenamefont [1]{#1}%
\providecommand \href@noop [0]{\@secondoftwo}%
\providecommand \href [0]{\begingroup \@sanitize@url \@href}%
\providecommand \@href[1]{\@@startlink{#1}\@@href}%
\providecommand \@@href[1]{\endgroup#1\@@endlink}%
\providecommand \@sanitize@url [0]{\catcode `\\12\catcode `\$12\catcode
  `\&12\catcode `\#12\catcode `\^12\catcode `\_12\catcode `\%12\relax}%
\providecommand \@@startlink[1]{}%
\providecommand \@@endlink[0]{}%
\providecommand \url  [0]{\begingroup\@sanitize@url \@url }%
\providecommand \@url [1]{\endgroup\@href {#1}{\urlprefix }}%
\providecommand \urlprefix  [0]{URL }%
\providecommand \Eprint [0]{\href }%
\providecommand \doibase [0]{http://dx.doi.org/}%
\providecommand \selectlanguage [0]{\@gobble}%
\providecommand \bibinfo  [0]{\@secondoftwo}%
\providecommand \bibfield  [0]{\@secondoftwo}%
\providecommand \translation [1]{[#1]}%
\providecommand \BibitemOpen [0]{}%
\providecommand \bibitemStop [0]{}%
\providecommand \bibitemNoStop [0]{.\EOS\space}%
\providecommand \EOS [0]{\spacefactor3000\relax}%
\providecommand \BibitemShut  [1]{\csname bibitem#1\endcsname}%
\let\auto@bib@innerbib\@empty
\bibitem [{\citenamefont {Freedman}(1974)}]{Freedman:1973yd}%
  \BibitemOpen
  \bibfield  {author} {\bibinfo {author} {\bibfnamefont {D.~Z.}\ \bibnamefont
  {Freedman}},\ }\href {\doibase 10.1103/PhysRevD.9.1389} {\bibfield  {journal}
  {\bibinfo  {journal} {Phys.\ Rev.\ D}\ }\textbf {\bibinfo {volume} {9}},\
  \bibinfo {pages} {1389} (\bibinfo {year} {1974})}\BibitemShut {NoStop}%
\bibitem [{\citenamefont {Kopeliovich}\ and\ \citenamefont
  {Frankfurt}(1974)}]{Kopeliovich:1974mv}%
  \BibitemOpen
  \bibfield  {author} {\bibinfo {author} {\bibfnamefont {V.}~\bibnamefont
  {Kopeliovich}}\ and\ \bibinfo {author} {\bibfnamefont {L.}~\bibnamefont
  {Frankfurt}},\ }\href@noop {} {\bibfield  {journal} {\bibinfo  {journal}
  {JETP Lett.}\ }\textbf {\bibinfo {volume} {19}},\ \bibinfo {pages} {145}
  (\bibinfo {year} {1974})}\BibitemShut {NoStop}%
\bibitem [{\citenamefont {Drukier}\ and\ \citenamefont
  {Stodolsky}(1984)}]{Drukier:1983gj}%
  \BibitemOpen
  \bibfield  {author} {\bibinfo {author} {\bibfnamefont {A.}~\bibnamefont
  {Drukier}}\ and\ \bibinfo {author} {\bibfnamefont {L.}~\bibnamefont
  {Stodolsky}},\ }\href {\doibase 10.1103/PhysRevD.30.2295} {\ ,\ \bibinfo
  {pages} {395} (\bibinfo {year} {1984})}\BibitemShut {NoStop}%
\bibitem [{\citenamefont {Cabrera}\ \emph {et~al.}(1985)\citenamefont
  {Cabrera}, \citenamefont {Krauss},\ and\ \citenamefont
  {Wilczek}}]{PhysRevLett.55.25}%
  \BibitemOpen
  \bibfield  {author} {\bibinfo {author} {\bibfnamefont {B.}~\bibnamefont
  {Cabrera}}, \bibinfo {author} {\bibfnamefont {L.~M.}\ \bibnamefont {Krauss}},
  \ and\ \bibinfo {author} {\bibfnamefont {F.}~\bibnamefont {Wilczek}},\ }\href
  {\doibase 10.1103/PhysRevLett.55.25} {\bibfield  {journal} {\bibinfo
  {journal} {Phys. Rev. Lett.}\ }\textbf {\bibinfo {volume} {55}},\ \bibinfo
  {pages} {25} (\bibinfo {year} {1985})}\BibitemShut {NoStop}%
\bibitem [{\citenamefont {Goodman}\ and\ \citenamefont
  {Witten}(1985)}]{Goodman:1984dc}%
  \BibitemOpen
  \bibfield  {author} {\bibinfo {author} {\bibfnamefont {M.~W.}\ \bibnamefont
  {Goodman}}\ and\ \bibinfo {author} {\bibfnamefont {E.}~\bibnamefont
  {Witten}},\ }\href {\doibase 10.1103/PhysRevD.31.3059} {\bibfield  {journal}
  {\bibinfo  {journal} {Phys.\ Rev.\ D}\ }\textbf {\bibinfo {volume} {31}},\
  \bibinfo {pages} {3059} (\bibinfo {year} {1985})}\BibitemShut {NoStop}%
\bibitem [{\citenamefont {Strigari}(2009)}]{Strigari:2009bq}%
  \BibitemOpen
  \bibfield  {author} {\bibinfo {author} {\bibfnamefont {L.~E.}\ \bibnamefont
  {Strigari}},\ }\href {\doibase 10.1088/1367-2630/11/10/105011} {\bibfield
  {journal} {\bibinfo  {journal} {New J.\ Phys.}\ }\textbf {\bibinfo {volume}
  {11}},\ \bibinfo {pages} {105011} (\bibinfo {year} {2009})},\ \Eprint
  {http://arxiv.org/abs/0903.3630} {arXiv:0903.3630 [astro-ph.CO]} \BibitemShut
  {NoStop}%
\bibitem [{\citenamefont {Billard}\ \emph {et~al.}(2014)\citenamefont
  {Billard}, \citenamefont {Strigari},\ and\ \citenamefont
  {Figueroa-Feliciano}}]{Billard:2013qya}%
  \BibitemOpen
  \bibfield  {author} {\bibinfo {author} {\bibfnamefont {J.}~\bibnamefont
  {Billard}}, \bibinfo {author} {\bibfnamefont {L.}~\bibnamefont {Strigari}}, \
  and\ \bibinfo {author} {\bibfnamefont {E.}~\bibnamefont
  {Figueroa-Feliciano}},\ }\href {\doibase 10.1103/PhysRevD.89.023524}
  {\bibfield  {journal} {\bibinfo  {journal} {Phys. Rev.}\ }\textbf {\bibinfo
  {volume} {D89}},\ \bibinfo {pages} {023524} (\bibinfo {year} {2014})},\
  \Eprint {http://arxiv.org/abs/1307.5458} {arXiv:1307.5458 [hep-ph]}
  \BibitemShut {NoStop}%
\bibitem [{\citenamefont {Monroe}\ and\ \citenamefont
  {Fisher}(2007)}]{Monroe:2007xp}%
  \BibitemOpen
  \bibfield  {author} {\bibinfo {author} {\bibfnamefont {J.}~\bibnamefont
  {Monroe}}\ and\ \bibinfo {author} {\bibfnamefont {P.}~\bibnamefont
  {Fisher}},\ }\href {\doibase 10.1103/PhysRevD.76.033007} {\bibfield
  {journal} {\bibinfo  {journal} {Phys. Rev.}\ }\textbf {\bibinfo {volume}
  {D76}},\ \bibinfo {pages} {033007} (\bibinfo {year} {2007})},\ \Eprint
  {http://arxiv.org/abs/0706.3019} {arXiv:0706.3019 [astro-ph]} \BibitemShut
  {NoStop}%
\bibitem [{\citenamefont {Battaglieri}\ \emph {et~al.}(2017)\citenamefont
  {Battaglieri} \emph {et~al.}}]{Battaglieri:2017aum}%
  \BibitemOpen
  \bibfield  {author} {\bibinfo {author} {\bibfnamefont {M.}~\bibnamefont
  {Battaglieri}} \emph {et~al.},\ }\href@noop {} {\  (\bibinfo {year}
  {2017})},\ \Eprint {http://arxiv.org/abs/1707.04591} {arXiv:1707.04591
  [hep-ph]} \BibitemShut {NoStop}%
\bibitem [{\citenamefont {Akerib}\ \emph {et~al.}(2017)\citenamefont {Akerib}
  \emph {et~al.}}]{Akerib:2016vxi}%
  \BibitemOpen
  \bibfield  {author} {\bibinfo {author} {\bibfnamefont {D.~S.}\ \bibnamefont
  {Akerib}} \emph {et~al.} (\bibinfo {collaboration} {LUX}),\ }\href {\doibase
  10.1103/PhysRevLett.118.021303} {\bibfield  {journal} {\bibinfo  {journal}
  {Phys. Rev. Lett.}\ }\textbf {\bibinfo {volume} {118}},\ \bibinfo {pages}
  {021303} (\bibinfo {year} {2017})},\ \Eprint
  {http://arxiv.org/abs/1608.07648} {arXiv:1608.07648 [astro-ph.CO]}
  \BibitemShut {NoStop}%
\bibitem [{\citenamefont {Cui}\ \emph {et~al.}(2017)\citenamefont {Cui} \emph
  {et~al.}}]{Cui:2017nnn}%
  \BibitemOpen
  \bibfield  {author} {\bibinfo {author} {\bibfnamefont {X.}~\bibnamefont
  {Cui}} \emph {et~al.} (\bibinfo {collaboration} {PandaX-II}),\ }\href
  {\doibase 10.1103/PhysRevLett.119.181302} {\bibfield  {journal} {\bibinfo
  {journal} {Phys.\ Rev.\ Lett.}\ }\textbf {\bibinfo {volume} {119}},\ \bibinfo
  {pages} {181302} (\bibinfo {year} {2017})},\ \Eprint
  {http://arxiv.org/abs/1708.06917} {arXiv:1708.06917 [astro-ph.CO]}
  \BibitemShut {NoStop}%
\bibitem [{\citenamefont {Aprile}\ \emph {et~al.}(2018)\citenamefont {Aprile}
  \emph {et~al.}}]{Aprile:2018dbl}%
  \BibitemOpen
  \bibfield  {author} {\bibinfo {author} {\bibfnamefont {E.}~\bibnamefont
  {Aprile}} \emph {et~al.} (\bibinfo {collaboration} {XENON}),\ }\href
  {\doibase 10.1103/PhysRevLett.121.111302} {\bibfield  {journal} {\bibinfo
  {journal} {Phys.\ Rev.\ Lett.}\ }\textbf {\bibinfo {volume} {121}},\ \bibinfo
  {pages} {111302} (\bibinfo {year} {2018})},\ \Eprint
  {http://arxiv.org/abs/1805.12562} {arXiv:1805.12562 [astro-ph.CO]}
  \BibitemShut {NoStop}%
\bibitem [{\citenamefont {Ajaj}\ \emph {et~al.}(2019)\citenamefont {Ajaj} \emph
  {et~al.}}]{Ajaj:2019imk}%
  \BibitemOpen
  \bibfield  {author} {\bibinfo {author} {\bibfnamefont {R.}~\bibnamefont
  {Ajaj}} \emph {et~al.} (\bibinfo {collaboration} {DEAP}),\ }\href {\doibase
  10.1103/PhysRevD.100.022004} {\bibfield  {journal} {\bibinfo  {journal}
  {Phys.\ Rev.\ D}\ }\textbf {\bibinfo {volume} {100}},\ \bibinfo {pages}
  {022004} (\bibinfo {year} {2019})},\ \Eprint
  {http://arxiv.org/abs/1902.04048} {arXiv:1902.04048 [astro-ph.CO]}
  \BibitemShut {NoStop}%
\bibitem [{\citenamefont {Agnes}\ \emph {et~al.}(2018)\citenamefont {Agnes}
  \emph {et~al.}}]{Agnes:2018fwg}%
  \BibitemOpen
  \bibfield  {author} {\bibinfo {author} {\bibfnamefont {P.}~\bibnamefont
  {Agnes}} \emph {et~al.} (\bibinfo {collaboration} {DarkSide}),\ }\href
  {\doibase 10.1103/PhysRevD.98.102006} {\bibfield  {journal} {\bibinfo
  {journal} {Phys.\ Rev.\ D}\ }\textbf {\bibinfo {volume} {98}},\ \bibinfo
  {pages} {102006} (\bibinfo {year} {2018})},\ \Eprint
  {http://arxiv.org/abs/1802.07198} {arXiv:1802.07198 [astro-ph.CO]}
  \BibitemShut {NoStop}%
\bibitem [{\citenamefont {Aprile}\ \emph {et~al.}(2016)\citenamefont {Aprile}
  \emph {et~al.}}]{Aprile:2015uzo}%
  \BibitemOpen
  \bibfield  {author} {\bibinfo {author} {\bibfnamefont {E.}~\bibnamefont
  {Aprile}} \emph {et~al.} (\bibinfo {collaboration} {XENON}),\ }\href
  {\doibase 10.1088/1475-7516/2016/04/027} {\bibfield  {journal} {\bibinfo
  {journal} {JCAP}\ }\textbf {\bibinfo {volume} {1604}},\ \bibinfo {pages}
  {027} (\bibinfo {year} {2016})},\ \Eprint {http://arxiv.org/abs/1512.07501}
  {arXiv:1512.07501 [physics.ins-det]} \BibitemShut {NoStop}%
\bibitem [{\citenamefont {Akerib}\ \emph {et~al.}(2020)\citenamefont {Akerib}
  \emph {et~al.}}]{Akerib:2018dfk}%
  \BibitemOpen
  \bibfield  {author} {\bibinfo {author} {\bibfnamefont {D.~S.}\ \bibnamefont
  {Akerib}} \emph {et~al.} (\bibinfo {collaboration} {LUX-ZEPLIN}),\ }\href
  {\doibase 10.1103/PhysRevD.101.052002} {\bibfield  {journal} {\bibinfo
  {journal} {Phys. Rev.}\ }\textbf {\bibinfo {volume} {D101}},\ \bibinfo
  {pages} {052002} (\bibinfo {year} {2020})},\ \Eprint
  {http://arxiv.org/abs/1802.06039} {arXiv:1802.06039 [astro-ph.IM]}
  \BibitemShut {NoStop}%
\bibitem [{\citenamefont {Aalbers}\ \emph {et~al.}(2016)\citenamefont {Aalbers}
  \emph {et~al.}}]{Aalbers:2016jon}%
  \BibitemOpen
  \bibfield  {author} {\bibinfo {author} {\bibfnamefont {J.}~\bibnamefont
  {Aalbers}} \emph {et~al.} (\bibinfo {collaboration} {DARWIN}),\ }\href
  {\doibase 10.1088/1475-7516/2016/11/017} {\bibfield  {journal} {\bibinfo
  {journal} {JCAP}\ }\textbf {\bibinfo {volume} {1611}},\ \bibinfo {pages}
  {017} (\bibinfo {year} {2016})},\ \Eprint {http://arxiv.org/abs/1606.07001}
  {arXiv:1606.07001 [astro-ph.IM]} \BibitemShut {NoStop}%
\bibitem [{\citenamefont {Aalseth}\ \emph {et~al.}(2018)\citenamefont {Aalseth}
  \emph {et~al.}}]{Aalseth:2017fik}%
  \BibitemOpen
  \bibfield  {author} {\bibinfo {author} {\bibfnamefont {C.~E.}\ \bibnamefont
  {Aalseth}} \emph {et~al.},\ }\href {\doibase 10.1140/epjp/i2018-11973-4}
  {\bibfield  {journal} {\bibinfo  {journal} {Eur. Phys. J. Plus}\ }\textbf
  {\bibinfo {volume} {133}},\ \bibinfo {pages} {131} (\bibinfo {year}
  {2018})},\ \Eprint {http://arxiv.org/abs/1707.08145} {arXiv:1707.08145
  [physics.ins-det]} \BibitemShut {NoStop}%
\bibitem [{\citenamefont {Akimov}\ \emph {et~al.}(2017)\citenamefont {Akimov}
  \emph {et~al.}}]{Akimov:2017ade}%
  \BibitemOpen
  \bibfield  {author} {\bibinfo {author} {\bibfnamefont {D.}~\bibnamefont
  {Akimov}} \emph {et~al.} (\bibinfo {collaboration} {COHERENT}),\ }\href
  {\doibase 10.1126/science.aao0990} {\bibfield  {journal} {\bibinfo  {journal}
  {Science}\ }\textbf {\bibinfo {volume} {357}},\ \bibinfo {pages} {1123}
  (\bibinfo {year} {2017})},\ \Eprint {http://arxiv.org/abs/1708.01294}
  {arXiv:1708.01294 [nucl-ex]} \BibitemShut {NoStop}%
\bibitem [{\citenamefont {Akimov}\ \emph {et~al.}(2020)\citenamefont {Akimov}
  \emph {et~al.}}]{Akimov:2020pdx}%
  \BibitemOpen
  \bibfield  {author} {\bibinfo {author} {\bibfnamefont {D.}~\bibnamefont
  {Akimov}} \emph {et~al.} (\bibinfo {collaboration} {COHERENT}),\ }\href@noop
  {} {\  (\bibinfo {year} {2020})},\ \Eprint {http://arxiv.org/abs/2003.10630}
  {arXiv:2003.10630 [nucl-ex]} \BibitemShut {NoStop}%
\bibitem [{\citenamefont {Angloher}\ \emph {et~al.}(2019)\citenamefont
  {Angloher} \emph {et~al.}}]{Angloher:2019flc}%
  \BibitemOpen
  \bibfield  {author} {\bibinfo {author} {\bibfnamefont {G.}~\bibnamefont
  {Angloher}} \emph {et~al.} (\bibinfo {collaboration} {NUCLEUS}),\ }\href
  {\doibase 10.1140/epjc/s10052-019-7454-4} {\bibfield  {journal} {\bibinfo
  {journal} {Eur. Phys. J.}\ }\textbf {\bibinfo {volume} {C79}},\ \bibinfo
  {pages} {1018} (\bibinfo {year} {2019})},\ \Eprint
  {http://arxiv.org/abs/1905.10258} {arXiv:1905.10258 [physics.ins-det]}
  \BibitemShut {NoStop}%
\bibitem [{\citenamefont {Hakenmüller}\ \emph {et~al.}(2019)\citenamefont
  {Hakenmüller} \emph {et~al.}}]{Hakenmuller:2019ecb}%
  \BibitemOpen
  \bibfield  {author} {\bibinfo {author} {\bibfnamefont {J.}~\bibnamefont
  {Hakenmüller}} \emph {et~al.},\ }\href {\doibase
  10.1140/epjc/s10052-019-7160-2} {\bibfield  {journal} {\bibinfo  {journal}
  {Eur.\ Phys.\ J.\ C}\ }\textbf {\bibinfo {volume} {79}},\ \bibinfo {pages}
  {699} (\bibinfo {year} {2019})},\ \Eprint {http://arxiv.org/abs/1903.09269}
  {arXiv:1903.09269 [physics.ins-det]} \BibitemShut {NoStop}%
\bibitem [{\citenamefont {Harnik}\ \emph {et~al.}(2012)\citenamefont {Harnik},
  \citenamefont {Kopp},\ and\ \citenamefont {Machado}}]{Harnik:2012ni}%
  \BibitemOpen
  \bibfield  {author} {\bibinfo {author} {\bibfnamefont {R.}~\bibnamefont
  {Harnik}}, \bibinfo {author} {\bibfnamefont {J.}~\bibnamefont {Kopp}}, \ and\
  \bibinfo {author} {\bibfnamefont {P.~A.~N.}\ \bibnamefont {Machado}},\ }\href
  {\doibase 10.1088/1475-7516/2012/07/026} {\bibfield  {journal} {\bibinfo
  {journal} {JCAP}\ }\textbf {\bibinfo {volume} {1207}},\ \bibinfo {pages}
  {026} (\bibinfo {year} {2012})},\ \Eprint {http://arxiv.org/abs/1202.6073}
  {arXiv:1202.6073 [hep-ph]} \BibitemShut {NoStop}%
\bibitem [{\citenamefont {Pospelov}\ and\ \citenamefont
  {Pradler}(2012)}]{Pospelov:2012gm}%
  \BibitemOpen
  \bibfield  {author} {\bibinfo {author} {\bibfnamefont {M.}~\bibnamefont
  {Pospelov}}\ and\ \bibinfo {author} {\bibfnamefont {J.}~\bibnamefont
  {Pradler}},\ }\href {\doibase 10.1103/PhysRevD.85.113016} {\bibfield
  {journal} {\bibinfo  {journal} {Phys.\ Rev.\ D}\ }\textbf {\bibinfo {volume}
  {85}},\ \bibinfo {pages} {113016} (\bibinfo {year} {2012})},\ \bibinfo {note}
  {[Erratum: Phys.Rev.D 88, 039904 (2013)]},\ \Eprint
  {http://arxiv.org/abs/1203.0545} {arXiv:1203.0545 [hep-ph]} \BibitemShut
  {NoStop}%
\bibitem [{\citenamefont {Billard}\ \emph {et~al.}(2015)\citenamefont
  {Billard}, \citenamefont {Strigari},\ and\ \citenamefont
  {Figueroa-Feliciano}}]{Billard:2014yka}%
  \BibitemOpen
  \bibfield  {author} {\bibinfo {author} {\bibfnamefont {J.}~\bibnamefont
  {Billard}}, \bibinfo {author} {\bibfnamefont {L.~E.}\ \bibnamefont
  {Strigari}}, \ and\ \bibinfo {author} {\bibfnamefont {E.}~\bibnamefont
  {Figueroa-Feliciano}},\ }\href {\doibase 10.1103/PhysRevD.91.095023}
  {\bibfield  {journal} {\bibinfo  {journal} {Phys. Rev.}\ }\textbf {\bibinfo
  {volume} {D91}},\ \bibinfo {pages} {095023} (\bibinfo {year} {2015})},\
  \Eprint {http://arxiv.org/abs/1409.0050} {arXiv:1409.0050 [astro-ph.CO]}
  \BibitemShut {NoStop}%
\bibitem [{\citenamefont {Dutta}\ \emph {et~al.}(2017)\citenamefont {Dutta},
  \citenamefont {Liao}, \citenamefont {Strigari},\ and\ \citenamefont
  {Walker}}]{Dutta:2017nht}%
  \BibitemOpen
  \bibfield  {author} {\bibinfo {author} {\bibfnamefont {B.}~\bibnamefont
  {Dutta}}, \bibinfo {author} {\bibfnamefont {S.}~\bibnamefont {Liao}},
  \bibinfo {author} {\bibfnamefont {L.~E.}\ \bibnamefont {Strigari}}, \ and\
  \bibinfo {author} {\bibfnamefont {J.~W.}\ \bibnamefont {Walker}},\ }\href
  {\doibase 10.1016/j.physletb.2017.08.031} {\bibfield  {journal} {\bibinfo
  {journal} {Phys. Lett.}\ }\textbf {\bibinfo {volume} {B773}},\ \bibinfo
  {pages} {242} (\bibinfo {year} {2017})},\ \Eprint
  {http://arxiv.org/abs/1705.00661} {arXiv:1705.00661 [hep-ph]} \BibitemShut
  {NoStop}%
\bibitem [{\citenamefont {Bertuzzo}\ \emph {et~al.}(2017)\citenamefont
  {Bertuzzo}, \citenamefont {Deppisch}, \citenamefont {Kulkarni}, \citenamefont
  {Perez~Gonzalez},\ and\ \citenamefont
  {Zukanovich~Funchal}}]{Bertuzzo:2017tuf}%
  \BibitemOpen
  \bibfield  {author} {\bibinfo {author} {\bibfnamefont {E.}~\bibnamefont
  {Bertuzzo}}, \bibinfo {author} {\bibfnamefont {F.~F.}\ \bibnamefont
  {Deppisch}}, \bibinfo {author} {\bibfnamefont {S.}~\bibnamefont {Kulkarni}},
  \bibinfo {author} {\bibfnamefont {Y.~F.}\ \bibnamefont {Perez~Gonzalez}}, \
  and\ \bibinfo {author} {\bibfnamefont {R.}~\bibnamefont
  {Zukanovich~Funchal}},\ }\href {\doibase 10.1007/JHEP04(2017)073} {\
  (\bibinfo {year} {2017}),\ 10.1007/JHEP04(2017)073},\ \Eprint
  {http://arxiv.org/abs/1701.07443} {arXiv:1701.07443 [hep-ph]} \BibitemShut
  {NoStop}%
\bibitem [{\citenamefont {Aristizabal~Sierra}\ \emph
  {et~al.}(2018)\citenamefont {Aristizabal~Sierra}, \citenamefont {Rojas},\
  and\ \citenamefont {Tytgat}}]{AristizabalSierra:2017joc}%
  \BibitemOpen
  \bibfield  {author} {\bibinfo {author} {\bibfnamefont {D.}~\bibnamefont
  {Aristizabal~Sierra}}, \bibinfo {author} {\bibfnamefont {N.}~\bibnamefont
  {Rojas}}, \ and\ \bibinfo {author} {\bibfnamefont {M.}~\bibnamefont
  {Tytgat}},\ }\href {\doibase 10.1007/JHEP03(2018)197} {\bibfield  {journal}
  {\bibinfo  {journal} {JHEP}\ }\textbf {\bibinfo {volume} {03}},\ \bibinfo
  {pages} {197} (\bibinfo {year} {2018})},\ \Eprint
  {http://arxiv.org/abs/1712.09667} {arXiv:1712.09667 [hep-ph]} \BibitemShut
  {NoStop}%
\bibitem [{\citenamefont {Shoemaker}\ and\ \citenamefont
  {Wyenberg}(2019)}]{Shoemaker:2018vii}%
  \BibitemOpen
  \bibfield  {author} {\bibinfo {author} {\bibfnamefont {I.~M.}\ \bibnamefont
  {Shoemaker}}\ and\ \bibinfo {author} {\bibfnamefont {J.}~\bibnamefont
  {Wyenberg}},\ }\href {\doibase 10.1103/PhysRevD.99.075010} {\bibfield
  {journal} {\bibinfo  {journal} {Phys. Rev. D}\ }\textbf {\bibinfo {volume}
  {99}},\ \bibinfo {pages} {075010} (\bibinfo {year} {2019})},\ \Eprint
  {http://arxiv.org/abs/1811.12435} {arXiv:1811.12435 [hep-ph]} \BibitemShut
  {NoStop}%
\bibitem [{\citenamefont {B\oe~hm}\ \emph {et~al.}(2019)\citenamefont
  {B\oe~hm}, \citenamefont {Cerdeño}, \citenamefont {Machado}, \citenamefont
  {Olivares-Del~Campo}, \citenamefont {Perdomo},\ and\ \citenamefont
  {Reid}}]{Boehm:2018sux}%
  \BibitemOpen
  \bibfield  {author} {\bibinfo {author} {\bibfnamefont {C.}~\bibnamefont
  {B\oe~hm}}, \bibinfo {author} {\bibfnamefont {D.}~\bibnamefont {Cerdeño}},
  \bibinfo {author} {\bibfnamefont {P.}~\bibnamefont {Machado}}, \bibinfo
  {author} {\bibfnamefont {A.}~\bibnamefont {Olivares-Del~Campo}}, \bibinfo
  {author} {\bibfnamefont {E.}~\bibnamefont {Perdomo}}, \ and\ \bibinfo
  {author} {\bibfnamefont {E.}~\bibnamefont {Reid}},\ }\href {\doibase
  10.1088/1475-7516/2019/01/043} {\bibfield  {journal} {\bibinfo  {journal}
  {JCAP}\ }\textbf {\bibinfo {volume} {01}},\ \bibinfo {pages} {043} (\bibinfo
  {year} {2019})},\ \Eprint {http://arxiv.org/abs/1809.06385} {arXiv:1809.06385
  [hep-ph]} \BibitemShut {NoStop}%
\bibitem [{\citenamefont {Gonzalez-Garcia}\ \emph {et~al.}(2018)\citenamefont
  {Gonzalez-Garcia}, \citenamefont {Maltoni}, \citenamefont {Perez-Gonzalez},\
  and\ \citenamefont {Zukanovich~Funchal}}]{Gonzalez-Garcia:2018dep}%
  \BibitemOpen
  \bibfield  {author} {\bibinfo {author} {\bibfnamefont {M.}~\bibnamefont
  {Gonzalez-Garcia}}, \bibinfo {author} {\bibfnamefont {M.}~\bibnamefont
  {Maltoni}}, \bibinfo {author} {\bibfnamefont {Y.~F.}\ \bibnamefont
  {Perez-Gonzalez}}, \ and\ \bibinfo {author} {\bibfnamefont {R.}~\bibnamefont
  {Zukanovich~Funchal}},\ }\href {\doibase 10.1007/JHEP07(2018)019} {\bibfield
  {journal} {\bibinfo  {journal} {JHEP}\ }\textbf {\bibinfo {volume} {07}},\
  \bibinfo {pages} {019} (\bibinfo {year} {2018})},\ \Eprint
  {http://arxiv.org/abs/1803.03650} {arXiv:1803.03650 [hep-ph]} \BibitemShut
  {NoStop}%
\bibitem [{\citenamefont {Aristizabal~Sierra}\ \emph
  {et~al.}(2019)\citenamefont {Aristizabal~Sierra}, \citenamefont {Dutta},
  \citenamefont {Liao},\ and\ \citenamefont
  {Strigari}}]{AristizabalSierra:2019ykk}%
  \BibitemOpen
  \bibfield  {author} {\bibinfo {author} {\bibfnamefont {D.}~\bibnamefont
  {Aristizabal~Sierra}}, \bibinfo {author} {\bibfnamefont {B.}~\bibnamefont
  {Dutta}}, \bibinfo {author} {\bibfnamefont {S.}~\bibnamefont {Liao}}, \ and\
  \bibinfo {author} {\bibfnamefont {L.~E.}\ \bibnamefont {Strigari}},\ }\href
  {\doibase 10.1007/JHEP12(2019)124} {\bibfield  {journal} {\bibinfo  {journal}
  {JHEP}\ }\textbf {\bibinfo {volume} {12}},\ \bibinfo {pages} {124} (\bibinfo
  {year} {2019})},\ \Eprint {http://arxiv.org/abs/1910.12437} {arXiv:1910.12437
  [hep-ph]} \BibitemShut {NoStop}%
\bibitem [{\citenamefont {Chao}\ \emph {et~al.}(2019)\citenamefont {Chao},
  \citenamefont {Jiang}, \citenamefont {Wang},\ and\ \citenamefont
  {Zhang}}]{Chao:2019pyh}%
  \BibitemOpen
  \bibfield  {author} {\bibinfo {author} {\bibfnamefont {W.}~\bibnamefont
  {Chao}}, \bibinfo {author} {\bibfnamefont {J.-G.}\ \bibnamefont {Jiang}},
  \bibinfo {author} {\bibfnamefont {X.}~\bibnamefont {Wang}}, \ and\ \bibinfo
  {author} {\bibfnamefont {X.-Y.}\ \bibnamefont {Zhang}},\ }\href {\doibase
  10.1088/1475-7516/2019/08/010} {\bibfield  {journal} {\bibinfo  {journal}
  {JCAP}\ }\textbf {\bibinfo {volume} {08}},\ \bibinfo {pages} {010} (\bibinfo
  {year} {2019})},\ \Eprint {http://arxiv.org/abs/1904.11214} {arXiv:1904.11214
  [hep-ph]} \BibitemShut {NoStop}%
\bibitem [{\citenamefont {Dutta}\ and\ \citenamefont
  {Strigari}(2019)}]{Dutta:2019oaj}%
  \BibitemOpen
  \bibfield  {author} {\bibinfo {author} {\bibfnamefont {B.}~\bibnamefont
  {Dutta}}\ and\ \bibinfo {author} {\bibfnamefont {L.~E.}\ \bibnamefont
  {Strigari}},\ }\href {\doibase 10.1146/annurev-nucl-101918-023450} {\bibfield
   {journal} {\bibinfo  {journal} {Ann.\ Rev.\ Nucl.\ Part.\ Sci.}\ }\textbf
  {\bibinfo {volume} {69}},\ \bibinfo {pages} {137} (\bibinfo {year} {2019})},\
  \Eprint {http://arxiv.org/abs/1901.08876} {arXiv:1901.08876 [hep-ph]}
  \BibitemShut {NoStop}%
\bibitem [{\citenamefont {Poulin}\ \emph {et~al.}(2016)\citenamefont {Poulin},
  \citenamefont {Serpico},\ and\ \citenamefont {Lesgourgues}}]{Poulin:2016nat}%
  \BibitemOpen
  \bibfield  {author} {\bibinfo {author} {\bibfnamefont {V.}~\bibnamefont
  {Poulin}}, \bibinfo {author} {\bibfnamefont {P.~D.}\ \bibnamefont {Serpico}},
  \ and\ \bibinfo {author} {\bibfnamefont {J.}~\bibnamefont {Lesgourgues}},\
  }\href {\doibase 10.1088/1475-7516/2016/08/036} {\bibfield  {journal}
  {\bibinfo  {journal} {JCAP}\ }\textbf {\bibinfo {volume} {1608}},\ \bibinfo
  {pages} {036} (\bibinfo {year} {2016})},\ \Eprint
  {http://arxiv.org/abs/1606.02073} {arXiv:1606.02073 [astro-ph.CO]}
  \BibitemShut {NoStop}%
\bibitem [{\citenamefont {Beacom}(2010)}]{Beacom:2010kk}%
  \BibitemOpen
  \bibfield  {author} {\bibinfo {author} {\bibfnamefont {J.~F.}\ \bibnamefont
  {Beacom}},\ }\href {\doibase 10.1146/annurev.nucl.010909.083331} {\bibfield
  {journal} {\bibinfo  {journal} {Ann. Rev. Nucl. Part. Sci.}\ }\textbf
  {\bibinfo {volume} {60}},\ \bibinfo {pages} {439} (\bibinfo {year} {2010})},\
  \Eprint {http://arxiv.org/abs/1004.3311} {arXiv:1004.3311 [astro-ph.HE]}
  \BibitemShut {NoStop}%
\bibitem [{\citenamefont {Beacom}\ and\ \citenamefont
  {Vagins}(2004)}]{Beacom:2003nk}%
  \BibitemOpen
  \bibfield  {author} {\bibinfo {author} {\bibfnamefont {J.~F.}\ \bibnamefont
  {Beacom}}\ and\ \bibinfo {author} {\bibfnamefont {M.~R.}\ \bibnamefont
  {Vagins}},\ }\href {\doibase 10.1103/PhysRevLett.93.171101} {\bibfield
  {journal} {\bibinfo  {journal} {Phys. Rev. Lett.}\ }\textbf {\bibinfo
  {volume} {93}},\ \bibinfo {pages} {171101} (\bibinfo {year} {2004})},\
  \Eprint {http://arxiv.org/abs/hep-ph/0309300} {arXiv:hep-ph/0309300 [hep-ph]}
  \BibitemShut {NoStop}%
\bibitem [{\citenamefont {Bays}\ \emph {et~al.}(2012)\citenamefont {Bays} \emph
  {et~al.}}]{Bays:2011si}%
  \BibitemOpen
  \bibfield  {author} {\bibinfo {author} {\bibfnamefont {K.}~\bibnamefont
  {Bays}} \emph {et~al.} (\bibinfo {collaboration} {Super-Kamiokande}),\ }\href
  {\doibase 10.1103/PhysRevD.85.052007} {\bibfield  {journal} {\bibinfo
  {journal} {Phys. Rev.}\ }\textbf {\bibinfo {volume} {D85}},\ \bibinfo {pages}
  {052007} (\bibinfo {year} {2012})},\ \Eprint {http://arxiv.org/abs/1111.5031}
  {arXiv:1111.5031 [hep-ex]} \BibitemShut {NoStop}%
\bibitem [{\citenamefont {Palomares-Ruiz}(2008)}]{PalomaresRuiz:2007ry}%
  \BibitemOpen
  \bibfield  {author} {\bibinfo {author} {\bibfnamefont {S.}~\bibnamefont
  {Palomares-Ruiz}},\ }\href {\doibase 10.1016/j.physletb.2008.05.040}
  {\bibfield  {journal} {\bibinfo  {journal} {Phys. Lett.}\ }\textbf {\bibinfo
  {volume} {B665}},\ \bibinfo {pages} {50} (\bibinfo {year} {2008})},\ \Eprint
  {http://arxiv.org/abs/0712.1937} {arXiv:0712.1937 [astro-ph]} \BibitemShut
  {NoStop}%
\bibitem [{\citenamefont {Garcia-Cely}\ and\ \citenamefont
  {Heeck}(2017)}]{Garcia-Cely:2017oco}%
  \BibitemOpen
  \bibfield  {author} {\bibinfo {author} {\bibfnamefont {C.}~\bibnamefont
  {Garcia-Cely}}\ and\ \bibinfo {author} {\bibfnamefont {J.}~\bibnamefont
  {Heeck}},\ }\href {\doibase 10.1007/JHEP05(2017)102} {\bibfield  {journal}
  {\bibinfo  {journal} {JHEP}\ }\textbf {\bibinfo {volume} {05}},\ \bibinfo
  {pages} {102} (\bibinfo {year} {2017})},\ \Eprint
  {http://arxiv.org/abs/1701.07209} {arXiv:1701.07209 [hep-ph]} \BibitemShut
  {NoStop}%
\bibitem [{\citenamefont {Palomares-Ruiz}\ and\ \citenamefont
  {Pascoli}(2008)}]{PalomaresRuiz:2007eu}%
  \BibitemOpen
  \bibfield  {author} {\bibinfo {author} {\bibfnamefont {S.}~\bibnamefont
  {Palomares-Ruiz}}\ and\ \bibinfo {author} {\bibfnamefont {S.}~\bibnamefont
  {Pascoli}},\ }\href {\doibase 10.1103/PhysRevD.77.025025} {\bibfield
  {journal} {\bibinfo  {journal} {Phys. Rev.}\ }\textbf {\bibinfo {volume}
  {D77}},\ \bibinfo {pages} {025025} (\bibinfo {year} {2008})},\ \Eprint
  {http://arxiv.org/abs/0710.5420} {arXiv:0710.5420 [astro-ph]} \BibitemShut
  {NoStop}%
\bibitem [{\citenamefont {Huang}\ and\ \citenamefont
  {Zhao}(2014)}]{Huang:2013xfa}%
  \BibitemOpen
  \bibfield  {author} {\bibinfo {author} {\bibfnamefont {J.}~\bibnamefont
  {Huang}}\ and\ \bibinfo {author} {\bibfnamefont {Y.}~\bibnamefont {Zhao}},\
  }\href {\doibase 10.1007/JHEP02(2014)077} {\bibfield  {journal} {\bibinfo
  {journal} {JHEP}\ }\textbf {\bibinfo {volume} {02}},\ \bibinfo {pages} {077}
  (\bibinfo {year} {2014})},\ \Eprint {http://arxiv.org/abs/1312.0011}
  {arXiv:1312.0011 [hep-ph]} \BibitemShut {NoStop}%
\bibitem [{\citenamefont {Agashe}\ \emph {et~al.}(2014)\citenamefont {Agashe},
  \citenamefont {Cui}, \citenamefont {Necib},\ and\ \citenamefont
  {Thaler}}]{Agashe:2014yua}%
  \BibitemOpen
  \bibfield  {author} {\bibinfo {author} {\bibfnamefont {K.}~\bibnamefont
  {Agashe}}, \bibinfo {author} {\bibfnamefont {Y.}~\bibnamefont {Cui}},
  \bibinfo {author} {\bibfnamefont {L.}~\bibnamefont {Necib}}, \ and\ \bibinfo
  {author} {\bibfnamefont {J.}~\bibnamefont {Thaler}},\ }\href {\doibase
  10.1088/1475-7516/2014/10/062} {\bibfield  {journal} {\bibinfo  {journal}
  {JCAP}\ }\textbf {\bibinfo {volume} {10}},\ \bibinfo {pages} {062} (\bibinfo
  {year} {2014})},\ \Eprint {http://arxiv.org/abs/1405.7370} {arXiv:1405.7370
  [hep-ph]} \BibitemShut {NoStop}%
\bibitem [{\citenamefont {Cui}\ \emph {et~al.}(2018)\citenamefont {Cui},
  \citenamefont {Pospelov},\ and\ \citenamefont {Pradler}}]{Cui:2017ytb}%
  \BibitemOpen
  \bibfield  {author} {\bibinfo {author} {\bibfnamefont {Y.}~\bibnamefont
  {Cui}}, \bibinfo {author} {\bibfnamefont {M.}~\bibnamefont {Pospelov}}, \
  and\ \bibinfo {author} {\bibfnamefont {J.}~\bibnamefont {Pradler}},\ }\href
  {\doibase 10.1103/PhysRevD.97.103004} {\bibfield  {journal} {\bibinfo
  {journal} {Phys. Rev.}\ }\textbf {\bibinfo {volume} {D97}},\ \bibinfo {pages}
  {103004} (\bibinfo {year} {2018})},\ \Eprint
  {http://arxiv.org/abs/1711.04531} {arXiv:1711.04531 [hep-ph]} \BibitemShut
  {NoStop}%
\bibitem [{\citenamefont {Navarro}\ \emph {et~al.}(1996)\citenamefont
  {Navarro}, \citenamefont {Frenk},\ and\ \citenamefont
  {White}}]{Navarro:1995iw}%
  \BibitemOpen
  \bibfield  {author} {\bibinfo {author} {\bibfnamefont {J.~F.}\ \bibnamefont
  {Navarro}}, \bibinfo {author} {\bibfnamefont {C.~S.}\ \bibnamefont {Frenk}},
  \ and\ \bibinfo {author} {\bibfnamefont {S.~D.~M.}\ \bibnamefont {White}},\
  }\href {\doibase 10.1086/177173} {\bibfield  {journal} {\bibinfo  {journal}
  {Astrophys. J.}\ }\textbf {\bibinfo {volume} {462}},\ \bibinfo {pages} {563}
  (\bibinfo {year} {1996})},\ \Eprint {http://arxiv.org/abs/astro-ph/9508025}
  {arXiv:astro-ph/9508025 [astro-ph]} \BibitemShut {NoStop}%
\bibitem [{\citenamefont {Aghanim}\ \emph {et~al.}(2018)\citenamefont {Aghanim}
  \emph {et~al.}}]{Aghanim:2018eyx}%
  \BibitemOpen
  \bibfield  {author} {\bibinfo {author} {\bibfnamefont {N.}~\bibnamefont
  {Aghanim}} \emph {et~al.} (\bibinfo {collaboration} {Planck}),\ }\href@noop
  {} {\  (\bibinfo {year} {2018})},\ \Eprint {http://arxiv.org/abs/1807.06209}
  {arXiv:1807.06209 [astro-ph.CO]} \BibitemShut {NoStop}%
\bibitem [{\citenamefont {Riess}\ \emph {et~al.}(2019)\citenamefont {Riess},
  \citenamefont {Casertano}, \citenamefont {Yuan}, \citenamefont {Macri},\ and\
  \citenamefont {Scolnic}}]{Riess:2019cxk}%
  \BibitemOpen
  \bibfield  {author} {\bibinfo {author} {\bibfnamefont {A.~G.}\ \bibnamefont
  {Riess}}, \bibinfo {author} {\bibfnamefont {S.}~\bibnamefont {Casertano}},
  \bibinfo {author} {\bibfnamefont {W.}~\bibnamefont {Yuan}}, \bibinfo {author}
  {\bibfnamefont {L.~M.}\ \bibnamefont {Macri}}, \ and\ \bibinfo {author}
  {\bibfnamefont {D.}~\bibnamefont {Scolnic}},\ }\href {\doibase
  10.3847/1538-4357/ab1422} {\bibfield  {journal} {\bibinfo  {journal}
  {Astrophys. J.}\ }\textbf {\bibinfo {volume} {876}},\ \bibinfo {pages} {85}
  (\bibinfo {year} {2019})},\ \Eprint {http://arxiv.org/abs/1903.07603}
  {arXiv:1903.07603 [astro-ph.CO]} \BibitemShut {NoStop}%
\bibitem [{\citenamefont {Pandey}\ \emph {et~al.}(2019)\citenamefont {Pandey},
  \citenamefont {Karwal},\ and\ \citenamefont {Das}}]{Pandey:2019plg}%
  \BibitemOpen
  \bibfield  {author} {\bibinfo {author} {\bibfnamefont {K.~L.}\ \bibnamefont
  {Pandey}}, \bibinfo {author} {\bibfnamefont {T.}~\bibnamefont {Karwal}}, \
  and\ \bibinfo {author} {\bibfnamefont {S.}~\bibnamefont {Das}},\ }\href@noop
  {} {\  (\bibinfo {year} {2019})},\ \Eprint {http://arxiv.org/abs/1902.10636}
  {arXiv:1902.10636 [astro-ph.CO]} \BibitemShut {NoStop}%
\bibitem [{\citenamefont {Vattis}\ \emph {et~al.}(2019)\citenamefont {Vattis},
  \citenamefont {Koushiappas},\ and\ \citenamefont {Loeb}}]{Vattis:2019efj}%
  \BibitemOpen
  \bibfield  {author} {\bibinfo {author} {\bibfnamefont {K.}~\bibnamefont
  {Vattis}}, \bibinfo {author} {\bibfnamefont {S.~M.}\ \bibnamefont
  {Koushiappas}}, \ and\ \bibinfo {author} {\bibfnamefont {A.}~\bibnamefont
  {Loeb}},\ }\href {\doibase 10.1103/PhysRevD.99.121302} {\bibfield  {journal}
  {\bibinfo  {journal} {Phys. Rev.}\ }\textbf {\bibinfo {volume} {D99}},\
  \bibinfo {pages} {121302} (\bibinfo {year} {2019})},\ \Eprint
  {http://arxiv.org/abs/1903.06220} {arXiv:1903.06220 [astro-ph.CO]}
  \BibitemShut {NoStop}%
\bibitem [{\citenamefont {Grevesse}\ and\ \citenamefont
  {Sauval}(1998)}]{Grevesse:1998bj}%
  \BibitemOpen
  \bibfield  {author} {\bibinfo {author} {\bibfnamefont {N.}~\bibnamefont
  {Grevesse}}\ and\ \bibinfo {author} {\bibfnamefont {A.~J.}\ \bibnamefont
  {Sauval}},\ }\href {\doibase 10.1023/A:1005161325181} {\bibfield  {journal}
  {\bibinfo  {journal} {Space Sci. Rev.}\ }\textbf {\bibinfo {volume} {85}},\
  \bibinfo {pages} {161} (\bibinfo {year} {1998})}\BibitemShut {NoStop}%
\bibitem [{\citenamefont {Asplund}\ \emph {et~al.}(2009)\citenamefont
  {Asplund}, \citenamefont {Grevesse}, \citenamefont {Sauval},\ and\
  \citenamefont {Scott}}]{Asplund:2009fu}%
  \BibitemOpen
  \bibfield  {author} {\bibinfo {author} {\bibfnamefont {M.}~\bibnamefont
  {Asplund}}, \bibinfo {author} {\bibfnamefont {N.}~\bibnamefont {Grevesse}},
  \bibinfo {author} {\bibfnamefont {A.~J.}\ \bibnamefont {Sauval}}, \ and\
  \bibinfo {author} {\bibfnamefont {P.}~\bibnamefont {Scott}},\ }\href
  {\doibase 10.1146/annurev.astro.46.060407.145222} {\bibfield  {journal}
  {\bibinfo  {journal} {Ann. Rev. Astron. Astrophys.}\ }\textbf {\bibinfo
  {volume} {47}},\ \bibinfo {pages} {481} (\bibinfo {year} {2009})},\ \Eprint
  {http://arxiv.org/abs/0909.0948} {arXiv:0909.0948 [astro-ph.SR]} \BibitemShut
  {NoStop}%
\bibitem [{\citenamefont {Vinyoles}\ \emph {et~al.}(2017)\citenamefont
  {Vinyoles}, \citenamefont {Serenelli}, \citenamefont {Villante},
  \citenamefont {Basu}, \citenamefont {Bergström}, \citenamefont
  {Gonzalez-Garcia}, \citenamefont {Maltoni}, \citenamefont {Peña-Garay},\
  and\ \citenamefont {Song}}]{Vinyoles:2016djt}%
  \BibitemOpen
  \bibfield  {author} {\bibinfo {author} {\bibfnamefont {N.}~\bibnamefont
  {Vinyoles}}, \bibinfo {author} {\bibfnamefont {A.~M.}\ \bibnamefont
  {Serenelli}}, \bibinfo {author} {\bibfnamefont {F.~L.}\ \bibnamefont
  {Villante}}, \bibinfo {author} {\bibfnamefont {S.}~\bibnamefont {Basu}},
  \bibinfo {author} {\bibfnamefont {J.}~\bibnamefont {Bergström}}, \bibinfo
  {author} {\bibfnamefont {M.~C.}\ \bibnamefont {Gonzalez-Garcia}}, \bibinfo
  {author} {\bibfnamefont {M.}~\bibnamefont {Maltoni}}, \bibinfo {author}
  {\bibfnamefont {C.}~\bibnamefont {Peña-Garay}}, \ and\ \bibinfo {author}
  {\bibfnamefont {N.}~\bibnamefont {Song}},\ }\href {\doibase
  10.3847/1538-4357/835/2/202} {\bibfield  {journal} {\bibinfo  {journal}
  {Astrophys. J.}\ }\textbf {\bibinfo {volume} {835}},\ \bibinfo {pages} {202}
  (\bibinfo {year} {2017})},\ \Eprint {http://arxiv.org/abs/1611.09867}
  {arXiv:1611.09867 [astro-ph.SR]} \BibitemShut {NoStop}%
\bibitem [{\citenamefont {Battistoni}\ \emph
  {et~al.}(2005{\natexlab{a}})\citenamefont {Battistoni}, \citenamefont
  {Ferrari}, \citenamefont {Montaruli},\ and\ \citenamefont
  {Sala}}]{BATTISTONI2005526}%
  \BibitemOpen
  \bibfield  {author} {\bibinfo {author} {\bibfnamefont {G.}~\bibnamefont
  {Battistoni}}, \bibinfo {author} {\bibfnamefont {A.}~\bibnamefont {Ferrari}},
  \bibinfo {author} {\bibfnamefont {T.}~\bibnamefont {Montaruli}}, \ and\
  \bibinfo {author} {\bibfnamefont {P.}~\bibnamefont {Sala}},\ }\href {\doibase
  https://doi.org/10.1016/j.astropartphys.2005.03.006} {\bibfield  {journal}
  {\bibinfo  {journal} {Astroparticle Physics}\ }\textbf {\bibinfo {volume}
  {23}},\ \bibinfo {pages} {526 } (\bibinfo {year}
  {2005}{\natexlab{a}})}\BibitemShut {NoStop}%
\bibitem [{\citenamefont {Totani}\ \emph {et~al.}(1996)\citenamefont {Totani},
  \citenamefont {Sato},\ and\ \citenamefont {Yoshii}}]{Totani:1995dw}%
  \BibitemOpen
  \bibfield  {author} {\bibinfo {author} {\bibfnamefont {T.}~\bibnamefont
  {Totani}}, \bibinfo {author} {\bibfnamefont {K.}~\bibnamefont {Sato}}, \ and\
  \bibinfo {author} {\bibfnamefont {Y.}~\bibnamefont {Yoshii}},\ }\href
  {\doibase 10.1086/176970} {\bibfield  {journal} {\bibinfo  {journal}
  {Astrophys. J.}\ }\textbf {\bibinfo {volume} {460}},\ \bibinfo {pages} {303}
  (\bibinfo {year} {1996})},\ \Eprint {http://arxiv.org/abs/astro-ph/9509130}
  {arXiv:astro-ph/9509130} \BibitemShut {NoStop}%
\bibitem [{\citenamefont {Yuksel}\ \emph {et~al.}(2008)\citenamefont {Yuksel},
  \citenamefont {Kistler}, \citenamefont {Beacom},\ and\ \citenamefont
  {Hopkins}}]{Yuksel:2008cu}%
  \BibitemOpen
  \bibfield  {author} {\bibinfo {author} {\bibfnamefont {H.}~\bibnamefont
  {Yuksel}}, \bibinfo {author} {\bibfnamefont {M.~D.}\ \bibnamefont {Kistler}},
  \bibinfo {author} {\bibfnamefont {J.~F.}\ \bibnamefont {Beacom}}, \ and\
  \bibinfo {author} {\bibfnamefont {A.~M.}\ \bibnamefont {Hopkins}},\ }\href
  {\doibase 10.1086/591449} {\bibfield  {journal} {\bibinfo  {journal}
  {Astrophys. J.}\ }\textbf {\bibinfo {volume} {683}},\ \bibinfo {pages} {L5}
  (\bibinfo {year} {2008})},\ \Eprint {http://arxiv.org/abs/0804.4008}
  {arXiv:0804.4008 [astro-ph]} \BibitemShut {NoStop}%
\bibitem [{\citenamefont {Hartmann}\ and\ \citenamefont
  {Woosley}(1997)}]{HartmannD.H1997Tcsn}%
  \BibitemOpen
  \bibfield  {author} {\bibinfo {author} {\bibfnamefont {D.}~\bibnamefont
  {Hartmann}}\ and\ \bibinfo {author} {\bibfnamefont {S.}~\bibnamefont
  {Woosley}},\ }\href@noop {} {\bibfield  {journal} {\bibinfo  {journal}
  {Astroparticle Physics}\ }\textbf {\bibinfo {volume} {7}},\ \bibinfo {pages}
  {137} (\bibinfo {year} {1997})}\BibitemShut {NoStop}%
\bibitem [{\citenamefont {Gelmini}\ \emph {et~al.}(2018)\citenamefont
  {Gelmini}, \citenamefont {Takhistov},\ and\ \citenamefont
  {Witte}}]{Gelmini:2018ogy}%
  \BibitemOpen
  \bibfield  {author} {\bibinfo {author} {\bibfnamefont {G.~B.}\ \bibnamefont
  {Gelmini}}, \bibinfo {author} {\bibfnamefont {V.}~\bibnamefont {Takhistov}},
  \ and\ \bibinfo {author} {\bibfnamefont {S.~J.}\ \bibnamefont {Witte}},\
  }\href {\doibase 10.1088/1475-7516/2018/07/009,
  10.1088/1475-7516/2019/02/E02} {\bibfield  {journal} {\bibinfo  {journal}
  {JCAP}\ }\textbf {\bibinfo {volume} {1807}},\ \bibinfo {pages} {009}
  (\bibinfo {year} {2018})},\ \bibinfo {note} {[Erratum: JCAP1902,E02(2019)]},\
  \Eprint {http://arxiv.org/abs/1804.01638} {arXiv:1804.01638 [hep-ph]}
  \BibitemShut {NoStop}%
\bibitem [{\citenamefont {Battistoni}\ \emph
  {et~al.}(2005{\natexlab{b}})\citenamefont {Battistoni}, \citenamefont
  {Ferrari}, \citenamefont {Montaruli},\ and\ \citenamefont
  {Sala}}]{Battistoni:2005pd}%
  \BibitemOpen
  \bibfield  {author} {\bibinfo {author} {\bibfnamefont {G.}~\bibnamefont
  {Battistoni}}, \bibinfo {author} {\bibfnamefont {A.}~\bibnamefont {Ferrari}},
  \bibinfo {author} {\bibfnamefont {T.}~\bibnamefont {Montaruli}}, \ and\
  \bibinfo {author} {\bibfnamefont {P.~R.}\ \bibnamefont {Sala}},\ }\href
  {\doibase 10.1016/j.astropartphys.2005.03.006} {\bibfield  {journal}
  {\bibinfo  {journal} {Astropart. Phys.}\ }\textbf {\bibinfo {volume} {23}},\
  \bibinfo {pages} {526} (\bibinfo {year} {2005}{\natexlab{b}})}\BibitemShut
  {NoStop}%
\bibitem [{\citenamefont {Horiuchi}\ \emph {et~al.}(2009)\citenamefont
  {Horiuchi}, \citenamefont {Beacom},\ and\ \citenamefont
  {Dwek}}]{Horiuchi:2008jz}%
  \BibitemOpen
  \bibfield  {author} {\bibinfo {author} {\bibfnamefont {S.}~\bibnamefont
  {Horiuchi}}, \bibinfo {author} {\bibfnamefont {J.~F.}\ \bibnamefont
  {Beacom}}, \ and\ \bibinfo {author} {\bibfnamefont {E.}~\bibnamefont
  {Dwek}},\ }\href {\doibase 10.1103/PhysRevD.79.083013} {\bibfield  {journal}
  {\bibinfo  {journal} {Phys. Rev.}\ }\textbf {\bibinfo {volume} {D79}},\
  \bibinfo {pages} {083013} (\bibinfo {year} {2009})},\ \Eprint
  {http://arxiv.org/abs/0812.3157} {arXiv:0812.3157 [astro-ph]} \BibitemShut
  {NoStop}%
\bibitem [{\citenamefont {Lewin}\ and\ \citenamefont
  {Smith}(1996)}]{Lewin:1995rx}%
  \BibitemOpen
  \bibfield  {author} {\bibinfo {author} {\bibfnamefont {J.~D.}\ \bibnamefont
  {Lewin}}\ and\ \bibinfo {author} {\bibfnamefont {P.~F.}\ \bibnamefont
  {Smith}},\ }\href {\doibase 10.1016/S0927-6505(96)00047-3} {\bibfield
  {journal} {\bibinfo  {journal} {Astropart. Phys.}\ }\textbf {\bibinfo
  {volume} {6}},\ \bibinfo {pages} {87} (\bibinfo {year} {1996})}\BibitemShut
  {NoStop}%
\bibitem [{\citenamefont {Pospelov}(2011)}]{Pospelov:2011ha}%
  \BibitemOpen
  \bibfield  {author} {\bibinfo {author} {\bibfnamefont {M.}~\bibnamefont
  {Pospelov}},\ }\href {\doibase 10.1103/PhysRevD.84.085008} {\bibfield
  {journal} {\bibinfo  {journal} {Phys.Rev.}\ }\textbf {\bibinfo {volume}
  {D84}},\ \bibinfo {pages} {085008} (\bibinfo {year} {2011})},\ \Eprint
  {http://arxiv.org/abs/1103.3261} {arXiv:1103.3261 [hep-ph]} \BibitemShut
  {NoStop}%
\bibitem [{\citenamefont {Pospelov}\ and\ \citenamefont
  {Pradler}(2014)}]{Pospelov:2013rha}%
  \BibitemOpen
  \bibfield  {author} {\bibinfo {author} {\bibfnamefont {M.}~\bibnamefont
  {Pospelov}}\ and\ \bibinfo {author} {\bibfnamefont {J.}~\bibnamefont
  {Pradler}},\ }\href {\doibase 10.1103/PhysRevD.89.055012} {\bibfield
  {journal} {\bibinfo  {journal} {Phys.Rev.}\ }\textbf {\bibinfo {volume}
  {D89}},\ \bibinfo {pages} {055012} (\bibinfo {year} {2014})},\ \Eprint
  {http://arxiv.org/abs/1311.5764} {arXiv:1311.5764 [hep-ph]} \BibitemShut
  {NoStop}%
\bibitem [{\citenamefont {Batell}\ \emph {et~al.}(2014)\citenamefont {Batell},
  \citenamefont {deNiverville}, \citenamefont {McKeen}, \citenamefont
  {Pospelov},\ and\ \citenamefont {Ritz}}]{Batell:2014yra}%
  \BibitemOpen
  \bibfield  {author} {\bibinfo {author} {\bibfnamefont {B.}~\bibnamefont
  {Batell}}, \bibinfo {author} {\bibfnamefont {P.}~\bibnamefont
  {deNiverville}}, \bibinfo {author} {\bibfnamefont {D.}~\bibnamefont
  {McKeen}}, \bibinfo {author} {\bibfnamefont {M.}~\bibnamefont {Pospelov}}, \
  and\ \bibinfo {author} {\bibfnamefont {A.}~\bibnamefont {Ritz}},\ }\href@noop
  {} {\  (\bibinfo {year} {2014})},\ \Eprint {http://arxiv.org/abs/1405.7049}
  {arXiv:1405.7049 [hep-ph]} \BibitemShut {NoStop}%
\bibitem [{\citenamefont {Dror}\ \emph
  {et~al.}(2017{\natexlab{a}})\citenamefont {Dror}, \citenamefont {Lasenby},\
  and\ \citenamefont {Pospelov}}]{Dror:2017ehi}%
  \BibitemOpen
  \bibfield  {author} {\bibinfo {author} {\bibfnamefont {J.~A.}\ \bibnamefont
  {Dror}}, \bibinfo {author} {\bibfnamefont {R.}~\bibnamefont {Lasenby}}, \
  and\ \bibinfo {author} {\bibfnamefont {M.}~\bibnamefont {Pospelov}},\ }\href
  {\doibase 10.1103/PhysRevLett.119.141803} {\bibfield  {journal} {\bibinfo
  {journal} {Phys. Rev. Lett.}\ }\textbf {\bibinfo {volume} {119}},\ \bibinfo
  {pages} {141803} (\bibinfo {year} {2017}{\natexlab{a}})},\ \Eprint
  {http://arxiv.org/abs/1705.06726} {arXiv:1705.06726 [hep-ph]} \BibitemShut
  {NoStop}%
\bibitem [{\citenamefont {Dror}\ \emph
  {et~al.}(2017{\natexlab{b}})\citenamefont {Dror}, \citenamefont {Lasenby},\
  and\ \citenamefont {Pospelov}}]{Dror:2017nsg}%
  \BibitemOpen
  \bibfield  {author} {\bibinfo {author} {\bibfnamefont {J.~A.}\ \bibnamefont
  {Dror}}, \bibinfo {author} {\bibfnamefont {R.}~\bibnamefont {Lasenby}}, \
  and\ \bibinfo {author} {\bibfnamefont {M.}~\bibnamefont {Pospelov}},\ }\href
  {\doibase 10.1103/PhysRevD.96.075036} {\bibfield  {journal} {\bibinfo
  {journal} {Phys. Rev.}\ }\textbf {\bibinfo {volume} {D96}},\ \bibinfo {pages}
  {075036} (\bibinfo {year} {2017}{\natexlab{b}})},\ \Eprint
  {http://arxiv.org/abs/1707.01503} {arXiv:1707.01503 [hep-ph]} \BibitemShut
  {NoStop}%
\bibitem [{\citenamefont {Aguilar-Arevalo}\ \emph {et~al.}(2018)\citenamefont
  {Aguilar-Arevalo} \emph {et~al.}}]{Aguilar-Arevalo:2018wea}%
  \BibitemOpen
  \bibfield  {author} {\bibinfo {author} {\bibfnamefont {A.}~\bibnamefont
  {Aguilar-Arevalo}} \emph {et~al.} (\bibinfo {collaboration} {MiniBooNE DM}),\
  }\href {\doibase 10.1103/PhysRevD.98.112004} {\bibfield  {journal} {\bibinfo
  {journal} {Phys. Rev. D}\ }\textbf {\bibinfo {volume} {98}},\ \bibinfo
  {pages} {112004} (\bibinfo {year} {2018})},\ \Eprint
  {http://arxiv.org/abs/1807.06137} {arXiv:1807.06137 [hep-ex]} \BibitemShut
  {NoStop}%
\bibitem [{\citenamefont {Cowan}(1998)}]{cowan1998statistical}%
  \BibitemOpen
  \bibfield  {author} {\bibinfo {author} {\bibfnamefont {G.}~\bibnamefont
  {Cowan}},\ }\href@noop {} {\emph {\bibinfo {title} {Statistical data
  analysis}}}\ (\bibinfo  {publisher} {Oxford University Press, USA},\ \bibinfo
  {year} {1998})\BibitemShut {NoStop}%
\bibitem [{\citenamefont {Wilks}(1938)}]{wilks1938}%
  \BibitemOpen
  \bibfield  {author} {\bibinfo {author} {\bibfnamefont {S.~S.}\ \bibnamefont
  {Wilks}},\ }\href {\doibase 10.1214/aoms/1177732360} {\bibfield  {journal}
  {\bibinfo  {journal} {Ann. Math. Statist.}\ }\textbf {\bibinfo {volume}
  {9}},\ \bibinfo {pages} {60} (\bibinfo {year} {1938})}\BibitemShut {NoStop}%
\bibitem [{\citenamefont {Richard}\ \emph {et~al.}(2016)\citenamefont {Richard}
  \emph {et~al.}}]{Richard:2015aua}%
  \BibitemOpen
  \bibfield  {author} {\bibinfo {author} {\bibfnamefont {E.}~\bibnamefont
  {Richard}} \emph {et~al.} (\bibinfo {collaboration} {Super-Kamiokande}),\
  }\href {\doibase 10.1103/PhysRevD.94.052001} {\bibfield  {journal} {\bibinfo
  {journal} {Phys. Rev.}\ }\textbf {\bibinfo {volume} {D94}},\ \bibinfo {pages}
  {052001} (\bibinfo {year} {2016})},\ \Eprint
  {http://arxiv.org/abs/1510.08127} {arXiv:1510.08127 [hep-ex]} \BibitemShut
  {NoStop}%
\bibitem [{\citenamefont {Abe}\ \emph {et~al.}(2011)\citenamefont {Abe} \emph
  {et~al.}}]{Abe:2010hy}%
  \BibitemOpen
  \bibfield  {author} {\bibinfo {author} {\bibfnamefont {K.}~\bibnamefont
  {Abe}} \emph {et~al.} (\bibinfo {collaboration} {Super-Kamiokande}),\ }\href
  {\doibase 10.1103/PhysRevD.83.052010} {\bibfield  {journal} {\bibinfo
  {journal} {Phys. Rev.}\ }\textbf {\bibinfo {volume} {D83}},\ \bibinfo {pages}
  {052010} (\bibinfo {year} {2011})},\ \Eprint {http://arxiv.org/abs/1010.0118}
  {arXiv:1010.0118 [hep-ex]} \BibitemShut {NoStop}%
\bibitem [{\citenamefont {Bellini}\ \emph {et~al.}(2012)\citenamefont {Bellini}
  \emph {et~al.}}]{Bellini:2012kz}%
  \BibitemOpen
  \bibfield  {author} {\bibinfo {author} {\bibfnamefont {G.}~\bibnamefont
  {Bellini}} \emph {et~al.} (\bibinfo {collaboration} {Borexino}),\ }\href
  {\doibase 10.1103/PhysRevD.85.092003} {\bibfield  {journal} {\bibinfo
  {journal} {Phys. Rev.}\ }\textbf {\bibinfo {volume} {D85}},\ \bibinfo {pages}
  {092003} (\bibinfo {year} {2012})},\ \Eprint {http://arxiv.org/abs/1203.6258}
  {arXiv:1203.6258 [hep-ex]} \BibitemShut {NoStop}%
\end{thebibliography}%
\end{document}